# A Nonlinear Subspace Approach for Parametric Estimation of PDFs from Short Data Records with Application to Rayleigh Fading

**Ahmad A. Masoud**

Electrical Engineering Department, Center for communication systems and sensing, King Fahad University of Petroleum and Minerals, Dhahran Saudi Arabia , P.O. Box 287, 31261 (e-mail: masoud@kfupm.edu.sa)

**ABSTRACT** This paper tackles the issue of real-time parametric estimation of a wide class of probability density functions from limited datasets. This type of estimation addresses recent applications that require joint sensing and actuation. The suggested estimator operates in the nonlinear subspace that the parameter space of the distribution creates in the measurement sample space. This enables the estimator to embed *a priori* available information about the distribution in the computations to produce parameter estimates that are induced by signal components belonging only to the correct class of density functions being considered. It also enables it to nullify the effect of those components that do not belong to this class on the estimation process. The estimator can, with high accuracy, compute quickly the parameters of a wide class of probability density functions from short data records. The approach is developed and basic proofs of correctness are carried-out for the Rayleigh distribution, which is used to characterize wireless communication channels experiencing fast fading in heavily cluttered environments. Simulation results demonstrate the capabilities of the suggested procedure and the clear advantages it has over conventional norm-based estimation techniques. The results also show the ability of the suggested approach to estimate other density functions including the two-parameter lognormal distribution used to characterize shadowing in wireless communication.

**INDEX TERMS** Probability Density Estimation, Subspace Methods, Wireless Communication

### Symbols

| | |
|---|---|
| PDF: | Probability density function |
| N-D: | N-Dimensional |
| RV: | Random Variables |
| $P_X(x,\xi)$ : | Parameterized PDF |
| $\xi$ : | A vector containing L parameters of the PDF |
| $\boldsymbol{\Psi}(\xi)$ : | A vector constructed from N samples of the model PDF |
| $\hat{\Psi}_X$ : | A vector containing the histogram at the same sample location as $\boldsymbol{\Psi}(\xi)$ |
| K: | Number of random variable samples used to construct $\hat{\Psi}_X$ |
| N: | Number of PDF/Histogram samples |
| $\boldsymbol{\Lambda}$: | Parameter space |
| $\boldsymbol{S}$: | Sample space |
| $\boldsymbol{S}_{\Lambda}$: | Subspace which is formed in $\boldsymbol{S}$ as the image of $\boldsymbol{\Lambda}$ |
| $\Psi$x | Correct measurement due to the actual PDF of the random variable, |
| $\Psi e_t$ | Error component that belongs (tangent) to $\boldsymbol{S}_{\Lambda}$ |
| $\Psi$e | Error component that does not belong (normal) to $\boldsymbol{S}_{\Lambda}$ |
| D($\xi$) | A distance vector between the measurement and the estimate PDF |
| $J_{\xi}$ | Jacobian matrix with respect to the parameters of the PDF |
| $D_t$ | Component of D that is tangent to $\boldsymbol{S}_{\Lambda}$ |
| V($\xi$) | Lyapunov function constructed from D |
| $\dot{V}(\xi)$ | Time derivative of V($\xi$) |
| $\Xi$ | Set of points at which $\dot{V}(\xi)$ is zero |
| $\Omega$: | The minimum invariance set |
| $\Delta$ : | Set of equilibrium points of the dynamical system acting on $\xi$ |
| $M_{\boldsymbol{S}}$ | Measure of the sample space $\boldsymbol{S}$ |
| $M_{\boldsymbol{S}\Lambda}$ | Measure of the subspace ($\boldsymbol{S}_{\Lambda}$) |

## I. INTRODUCTION

Rich and accurate signal characterization is important for efficient utilization of processes. For example, wireless channel characteristics is a key factor in designing encoders for controlling the quality of data communication [1,2]. Properly actuating a valve to control the flow in a pipe requires characterizing the flow in that pipe to make actions coalesce with objectives.

It is common practice to characterize Signals using their measures. However, there are applications that require rich description of a process. A flow in a pipe is not just a number, which a metering device record. It's a random process that has a probability distribution whose parameters need to be estimated to better understand and predict the underlying phenomenon. Moreover, recent applications intertwine rich signal characterization and process actuation in a manner that most probably causes a change in the state of the process and the characteristics of the signal representing it. This requires timely and accurate update of the estimates.

The probability distribution function (PDF) of a signal is a strong characterizer that is able to produce many signal measures. Several techniques were suggested for the estimation of PDFs [3,4]. These techniques may be divided into two types: parametric estimators [5,6] and nonparametric estimators [7,8]. In many practical situations, the phenomenon concerned is well-examined and the type of PDF that describes it is accurately characterized up to a parameter ambiguity. The PDFs describing different types of communication channels are known beforehand [9, 10]. A communication channel in a heavily-cluttered environment experience fast fading that is best described by a Rayleigh PDF [11,12].

Communication-aware mobility is an important and fairly recent area in wireless communication [13,14,15] that requires on-line characterization of the PDF describing a wireless channel. In this area, a scout, first responder mobile agent that is equipped with a Tx/Rx antenna is required to access a confined highly cluttered space and wirelessly relay sensory data to a base station. The agent also receives motion servo data from that station. The nature of the situation causes considerable fading and shadowing effect (figure-1) in the wireless signal. This could, in addition to interrupting the data-feed to and from the base station, cause instability of motion [20]. The agent needs to online estimate the conditions of the wireless channel and make servo-level actuation decisions so that it will not lose stability or needlessly waste bandwidth. In such situations, a mobile agent has to wirelessly exchange causality packets with the network controller every few microseconds [16]. A causality packet contains a sensing action, a planning action, an actuation action and an estimation action. This leaves little time and little data for the computation of the channel PDF.

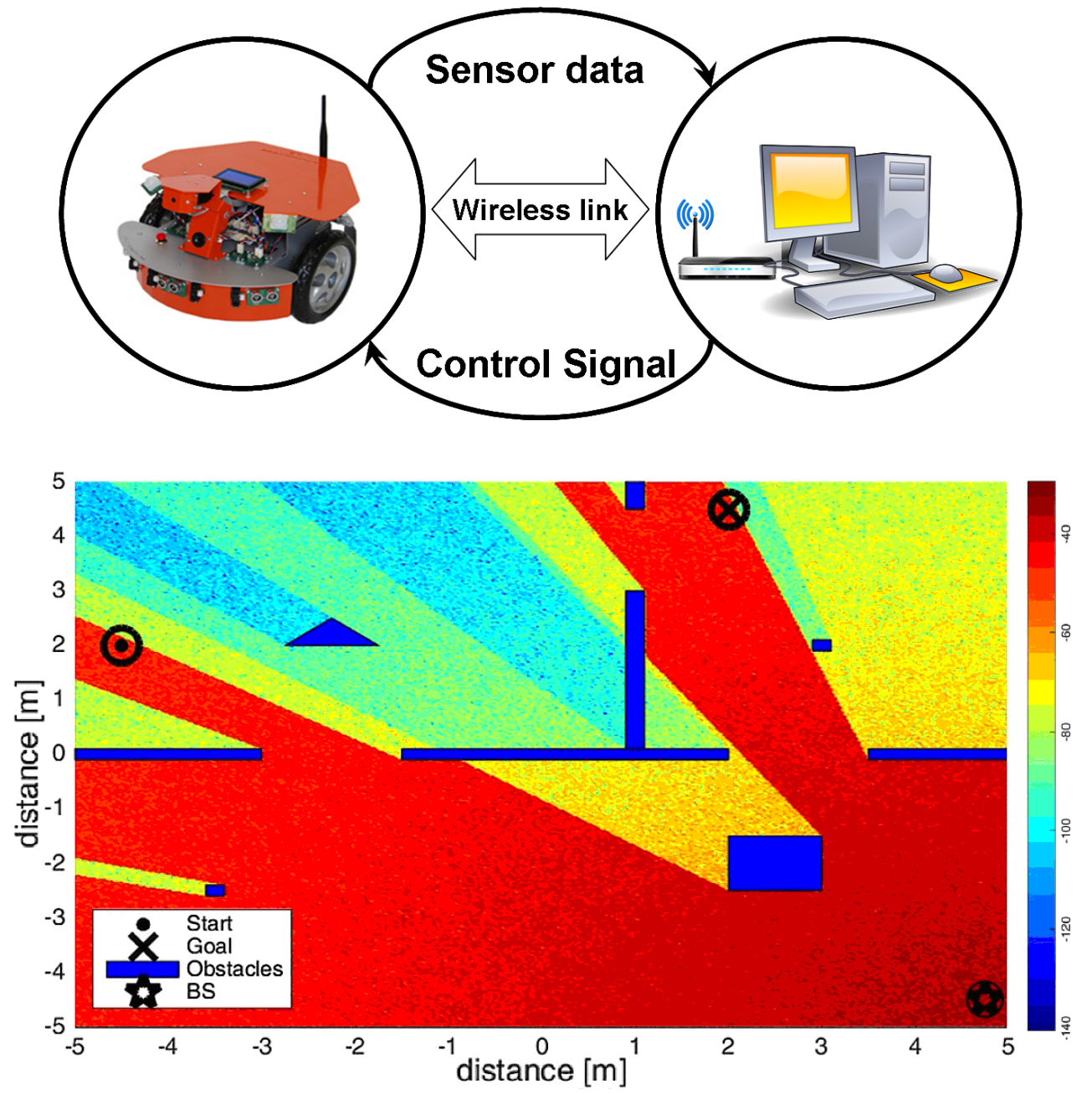

Figure-1: Wireless channel nature in cluttered confined environments

The Rayleigh distribution has long been considered to have important applications in many fields such as reliability theory, survival analysis, and, in particular, communication engineering [21,22]. A large number of techniques exist for the estimation of the parameter of this PDF. An unbiased closed form estimate of the Rayleigh parameter that utilizes the maximum likelihood approach is reported in [23]. The work in [24] derives closed form, Bayes-based parameter estimates that use Precautionary loss function, Entropy loss function and Loss function-L1. In [22] the best-unbiased closed form estimate in the Blackwell-Rao sense of the parameter is suggested. The authors in [28] provide a closed form value of the parameter estimate that is based on the moment method. There are also non-closed form estimators of the Rayleigh parameter that use Monte Carlo Expectation Maximization (MCEM) [25] and fuzzy logic [26]. Work is ongoing to develop generalizations of the Rayleigh distribution in order to enhance its utility. Examples of these generalizations are: the weighted Rayleigh distribution [27], the two-parameter Rayleigh distribution [28,29] and the generalized Rayleigh distribution [30].

Applications that employ joint actuation and estimation place stringent requirements on PDF computation. The estimation method should be able to cope with a variety of PDFs in a unified manner. While the Rayleigh distribution is of significant importance in wireless communication, other artifacts such as shadowing [31] and slow fading [32], which are best described by the lognormal distribution, may appear in the channel. Joint actuation and estimation also requires the estimators to be able to produce reliable and accurate estimates in real-time from small record of data. To the best of our knowledge, existing techniques for PDF estimation cannot simultaneously support all these requirements.

This paper suggests a novel parametric PDF estimation procedure that addresses effectively the issues of flexibility, speed and accuracy required by applications utilizing joint estimation and actuation. The estimation method uses model-based nonlinear subspace–restricted computations to filter noise from rough PDF estimates. These estimates are constructed from the histogram of the acquired random variable samples. Embedding information of the PDF model into the estimation process via a nonlinear subspace and restricting the estimates to lie in that subspace by using action reflection from parameter space to signal space enables the estimation process to resist high level of noise without the need to know the statistical nature of that noise. The generated estimate is optimum in the senses:

1- it guarantees that the estimate is extracted from data that is in strict conformity with the available a priori information that comes in the form of the PDF being considered
2- it nullifies the effect of invalid data that does not conform to the available model on the estimation process
3- it minimizes the error between the measurement and the set of all possible candidates that belong to the correct of PDF being estimated.

The proposed information-based constrained estimation significantly differs from the conventional norm-based estimation methods. These methods attempt direct minimization, in signal space, of an error norm representing the difference between the parameterized model PDF and the histogram-based measurement. As this paper demonstrates, obtaining parameter estimates in this manner can be problematic and highly unreliable when small data records of random variables are used.

This paper is organized as follows: section II develops the suggested nonlinear subspace estimator and explains its principle of operation. A realization of the estimator is provided in section III. The section also provides conditions for convergence and a proof that the Rayleigh PDF satisfies these conditions. Section IV demonstrates by simulation the capabilities of the estimator to compute the parameter of the Rayleigh PDF. It compares its performance to leading Rayleigh parameter estimators. It also demonstrates the impracticality of attempting to estimate the parameter by direct norm minimization. Section V discusses the fundamental differences between parameter estimation using direct norm minimization and estimation using the suggested nonlinear subspace method. Conclusions are placed in section VI.

## II. THE SUGGESTED ESTIMATOR

Consider a random variable $x$ with PDF $P_X(x,\xi)$ where $\xi$ is a vector containing the parameters of the distribution ($\xi = [\xi_1 \;\; .. \;\; \xi_L]^T$). Assume that the PDF is represented using its values at a set of samples $\{x_i,\ i=1,..N\}$. Those samples are selected in conformity with the sampling theorem so that the continuous PDF is uniquely determined from its discrete representation. The samples of the PDF are used to construct the parameterized N-D vector function (1)

$$\mathbf{\Psi}(\xi) = \begin{bmatrix} P_X(x_1,\xi) \\ : \\ P_X(x_N,\xi) \end{bmatrix}. \tag{1}$$

The suggested estimator treats the entries in this function as the coordinates of a sample space $\boldsymbol{S}$. The components of the parameter vector are also treated as the coordinates of the parameter space $\mathbf{\Lambda}$. A subspace ($\boldsymbol{S}_\Lambda$) is formed in $\boldsymbol{S}$ as the image of $\mathbf{\Lambda}$ under the vector transformation $\mathbf{\Psi}(\xi)$ (figure-2).

Let $\hat{\Psi}_X$ be a measurement of the PDF at the sample points $\{x_i,\ i=1,..N\}$. This measurement consists of three components (2)

$$\hat{\Psi}_X = \Psi_X + \Psi e_t + \Psi e_n \tag{2}$$

where $\Psi x$ is the correct measurement due to the actual PDF of the random variable, $\Psi e_t$ is the error component that belongs (tangent) to $\boldsymbol{S}_\Lambda$ (i.e. it belongs to the same type of PDFs being considered) and $\Psi e_n$ is the error component that is orthogonal to $\boldsymbol{S}_\Lambda$. If the estimation process is restricted to operate from within the parameter space with an equivalent effect that propagates to sample space, the effect of $\Psi e_n$ on the process is nullified and $\Psi e_t$ will be the only source of error. Blind estimation cannot eliminate this error since it represents a valid PDF. Only the use of side information will help in reducing the effect of this error on the estimate. Therefore, an optimal, blind processor will select the set of parameters that minimize the distance between the image of the PDF parameters' in sample space and the measurement which in our case translates to minimizing the projection of that distance on $\boldsymbol{S}_\Lambda$.

$$\min_{\xi} |D(\xi)| \qquad D(\xi) = \hat{\Psi} - \Psi(\xi) \tag{3}$$

## III. ESTIMATOR REALIZATION

This section provides a realization of the above estimator. In principle, the above approach can estimate any PDF provided that the distribution satisfies certain conditions. However, the focus here is on the Rayleigh PDF.

The measurement PDF ($\hat{\Psi}_X$) may be obtained from the histogram of the samples of the random variable. Histograms can accurately compute PDFs [17,18] provided a large record of RV samples is available. Since real time operation is required, $\hat{\Psi}_X$ is constructed using the largest sample record the situation permits. This will cause the measurement to be considerably noisy.

The minimization of D can only be carried-out by minimizing the component that belongs to $\boldsymbol{S}_\Lambda$ ($D_t$). This component results from the dot product between D and a complete set of vectors ($J_\xi$) that are tangent to $\boldsymbol{S}_\Lambda$

$$J_\xi = \frac{\partial \Psi(\xi)}{\partial \xi} = \begin{bmatrix} \frac{\partial P_X(x_1,\xi)}{\partial \xi_1} & \cdot\cdot & \frac{\partial P_X(x_1,\xi)}{\partial \xi_L} \\ \vdots & \cdot\cdot & \vdots \\ \frac{\partial P_X(x_N,\xi)}{\partial \xi_1} & \cdot\cdot & \frac{\partial P_X(x_N,\xi)}{\partial \xi_L} \end{bmatrix} \quad (4)$$

$$D_t = J_\xi^T(\xi) D(\xi)$$

By choosing $D_t$ as the action that controls the evolution of $\xi$, $D_t$ will converge to zero and $\xi$ will converge to a value that minimizes D

$$\dot{\xi} = J_\xi^T(\xi) D(\xi) \quad (5)$$

Figure-2 illustrates the estimation procedure.

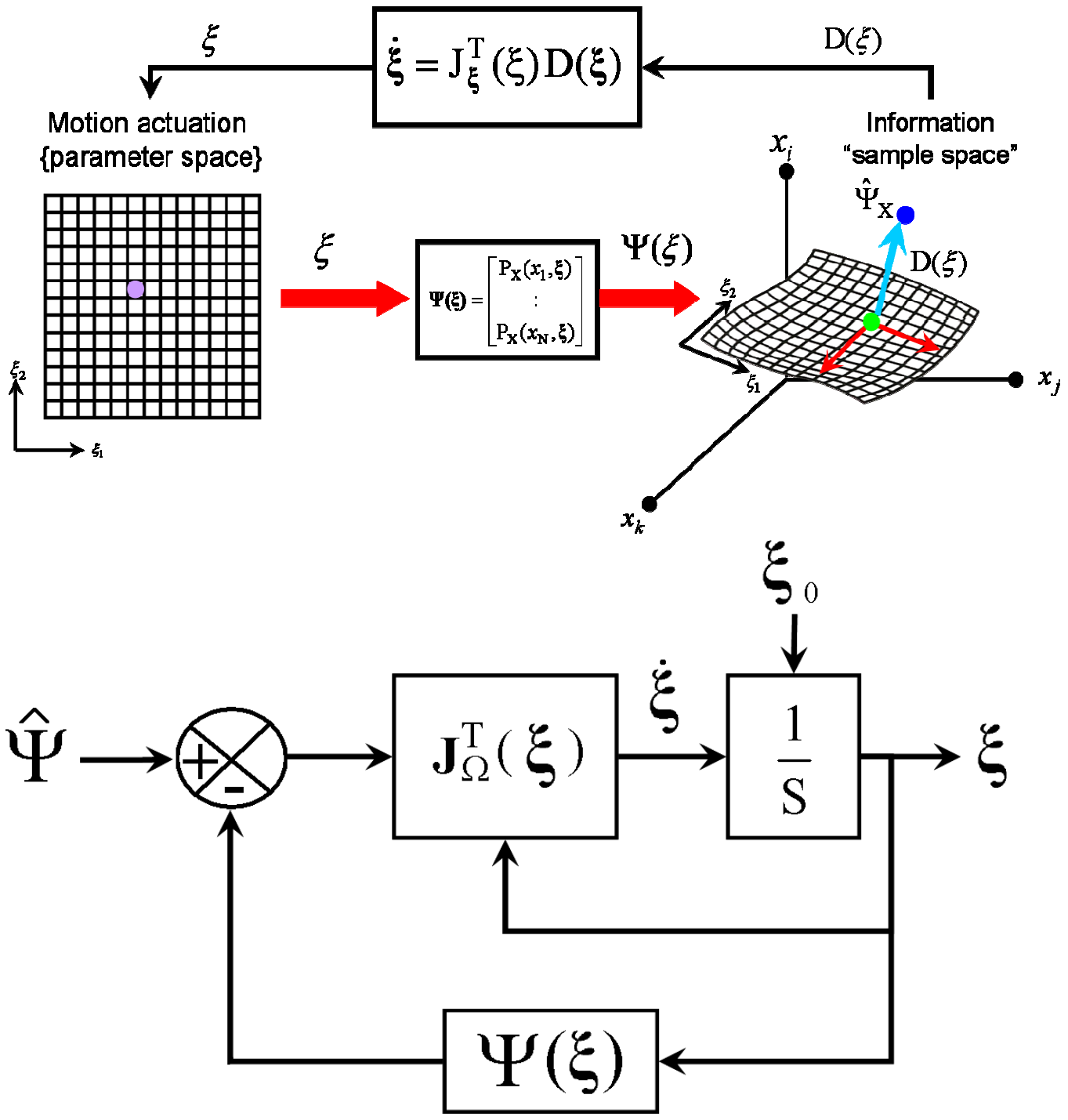

Figure-2: The Nonlinear Subspace Estimator

Ideally (i.e. the histogram measurement is noise-free), the first order dynamical system in (5) will globally asymptotically converge to the true value of the Rayleigh PDF parameter. The proposition below uses the LaSalle [19] invariance principle to prove the above.

Proposition-1: Consider a Rayleigh distributed random variable (6) with parameter $\sigma_o$.

$$P_X(x,\sigma_o) = \frac{x}{\sigma_o^2} \cdot \exp\left(-\frac{x^2}{2 \cdot \sigma_o^2}\right) \qquad x \geq 0 \quad (6)$$

If the measurement vector is constructed from properly chosen samples (7)

$$\hat{\Psi} = [P_X(x_1,\sigma_o) \ \ \ldots \ \ P_X(x_N,\sigma_o)]^T \quad (7)$$

then the dynamical system in (5) will converge to the true estimate from any initial choice of the parameter .

$$\lim_{t\to\infty} \xi(t) = \sigma_o \qquad \forall \xi(0) \quad (8)$$

Proof: consider the norm (Lyapunov Function)

$$V(\xi) = \frac{1}{2} D^T(\xi) D(\xi) \quad (9)$$

V is always positive and will equal zero if and only if D is zero. Since $\hat{\Psi}$ uniquely identifies the PDF, the convergence of D to zero implies the convergence of $\xi$ to $\sigma_o$. The time derivative of V is:

$$\dot{V}(\xi) = -D^T(\xi)\, \mathbf{J}(\xi)\, \dot{\xi} \quad (10)$$

If the derivative is selected as in (5) we have

$$\dot{V}(\xi) = -D^T(\xi)\, \mathbf{J}(\xi)\, \mathbf{J}^T(\xi)\, D(\xi) \quad (11)$$

The product of a matrix by its transpose, if not positive definite, is at least, positive semi-definite. In other words, the time derivative of the norm is negative semi-definite

$$\dot{V}(\xi) \leq 0 . \quad (12)$$

The set of $\xi$'s for which $\dot{V}(\xi) = 0$ must include among others the true value of the parameter (13)

$$\xi \in \Xi = \{\sigma_o \cup_i \xi_i, i = 1,..J\} \quad (13)$$

According to the LaSalle invariance principle, the dynamical system (5) will converge to the minimum invariance set ($\Omega$). To compute this set, first the set ($\Delta$) of $\xi$ for which system (5) is at equilibrium needs to be computed

$$\Delta = \{\xi : \mathbf{J}_\xi^T(\xi)\, D(\xi) = 0\} \quad (14)$$

For the Rayleigh distribution, the following approximation may be used

$$\mathbf{J}_\xi^T(\xi)\, D(\xi) \approx \alpha \cdot \int_0^\infty \left(\frac{d}{d\xi}\left(\frac{x}{\xi^2} \cdot \exp\left(-\frac{x^2}{2 \cdot \xi^2}\right)\right)\right)\left(\frac{x}{\sigma_o^2} \cdot \exp\left(-\frac{x^2}{2 \cdot \sigma_0^2} - \frac{x}{\xi^2} \cdot \exp\left(-\frac{x^2}{2 \cdot \xi^2}\right)\right)\right) dx \quad (15)$$

where α is a nonzero constant. By analyzing the above function one can show (figure-3) that (15) can only be zero at ξ equal to $\sigma_o$ ($\Delta=\{\sigma_o\}$). The minimum invariance set which ξ will converge to may be computed as

$$\Omega = \Xi \cap \Delta = \{\sigma_o\}\,. \qquad (16)$$

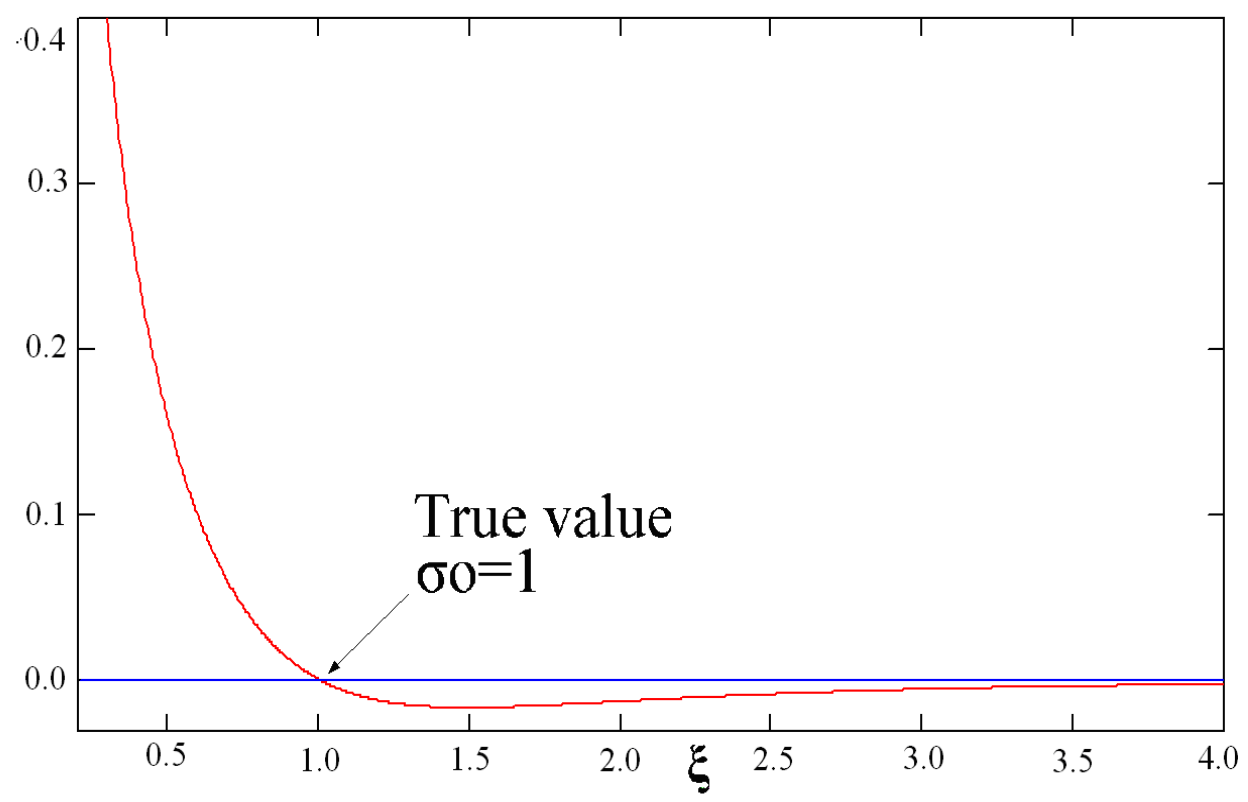

Figure-3: The dynamical system in (5) has one equilibrium point at the true estimate.

While convergence, in the ideal case is globally asymptotic, the presence of noise in the PDF measurement (histogram) causes the convergence to be local and to a finite set (17)

$$\lim_{t\to\infty} \xi(t) \in \left|\xi - \sigma_o\right| < \varepsilon \qquad \sigma_{mn} < \xi(0) < \sigma_{mx} \qquad (17)$$

where 0<ε<<1. As shown in the next section, the zone of convergence is by no means restrictive. It can accommodate an initial guess of the PDF parameter that is relatively far from the true value.

Convergence to a non-zero measure set (instead of a point), which is caused by noise, does not impose stringent limitations on the estimator's accuracy. The nature of the method makes it possible to easily and practically control the quality of the estimate. This may be concluded by considering two features of the estimation method. First, the true component of the PDF measurement maps with probability 1 to the $\boldsymbol{S}_\Lambda$ space while noise maps to the space with probability δ (figure-4). Since processing is carried-out fully in parameter space, any noise component of the measurement that lies in the space orthogonal to $\boldsymbol{S}_\Lambda$ will not affect the estimate.

The second issue has to do with the ability to control the error probability δ. Let $M_{\boldsymbol{S}}$ be a measure of the sample space $\boldsymbol{S}$ and $M_{\boldsymbol{S}\Lambda}$ be a measure of the subspace ($\boldsymbol{S}_\Lambda$). Notice that the dimensionality of $\boldsymbol{S}$ is equal to the number of histogram samples while the dimensionality of $\boldsymbol{S}_\Lambda$ is equal to the number of parameters and does not change with changing the number of histogram samples. As a result, $M_{\boldsymbol{S}}$ increases exponentially with the number of measurement (histogram) samples while $M_{\boldsymbol{S}\Lambda}$ stays the same. It is obvious that the probability of noise getting mapped into $\boldsymbol{S}_\Lambda$ and affecting the estimate considerably diminishes with increase in the number of histogram samples.

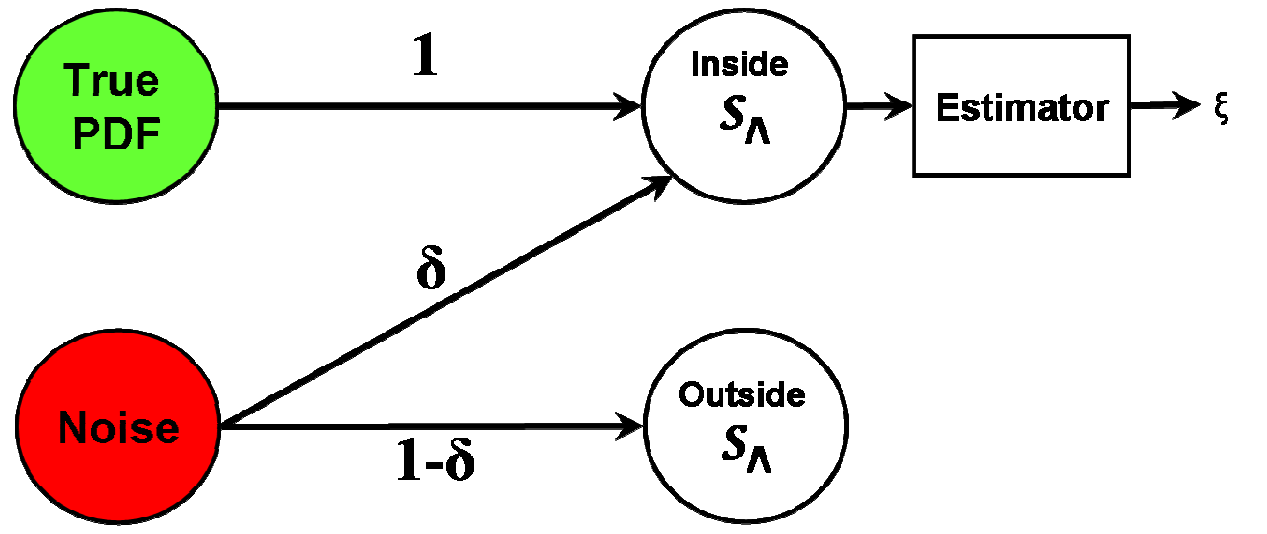

Figure-4: Mapping of the correct PDF and noise to the nonlinear subspace

The more random variables used in constructing a histogram, the less noisy is the histogram. The histogram noise is measured by the variance of the mean of the random variables used to construct a histogram sample. This variance is proportional to the inverse of the number of random variables used to construct a sample (K/N). If the number of random variable samples used to construct the histogram is kept constant, the histogram noise will increase with N.

The increase in noise caused by fixing the value of K and increasing N will be counteracted by an enhanced noise rejection capability of the subspace. Unless K is very small, the impact of noise on the quality of the estimate can be effectively managed by increasing N. The simulation results in the next section clearly demonstrate this behavior. Future work will focus on exploring this feature of the estimator mathematically.

## IV. RESULTS

This section demonstrates the capabilities of the suggested estimator. It examines the effect of the number of histogram samples and the number of random variable samples on the quality of the estimate. It also examines the sensitivity of the method to the value of the PDF's parameter, the convergence interval and the number of iterations needed for convergence. The convergence interval, which the initial condition of the parameter must lie in ($\xi(0) \in \{\sigma_{mn},\sigma_{mx}\}$), is determined experimentally.

The discrete realization of the nonlinear dynamical system in (5) does have a strong effect on the speed at which the estimate may be obtained and on its quality. Although an involved investigation of the dual statistical-dynamical nature of the estimator is needed in order to clearly understand its behavior, numerical investigation provides strong indicators about the estimator's capabilities.

In the first example a low number of random variable samples is used (K=50). The measurement of the PDF is constructed using 15 histogram samples only (N=15) and a low value for the parameter is selected ($\sigma_o$=1). The parameters of the trial are shown in table-1a. The estimator is initialized with three values, two close to the boundary of the convergence interval and one close to the true value.

| $\sigma_o$ | K | N | $\sigma_{mx}$ | $\sigma_{mn}$ |
|---|---|---|---|---|
| 1 | 50 | 15 | 2.5 | 0.1 |

Table-1a: Parameters of the estimator

| $\xi(0)$ | Iter | $\sigma_{est}$ | Err % |
|---|---|---|---|
| 2.5 | 65 | 0.986 | 1.4% |
| 0.5 | 35 | 0.986 | 1.4% |
| 0.15 | 43 | 0.986 | 1.4% |

Table-1b: Estimator results corresponding to parameters in table-1a

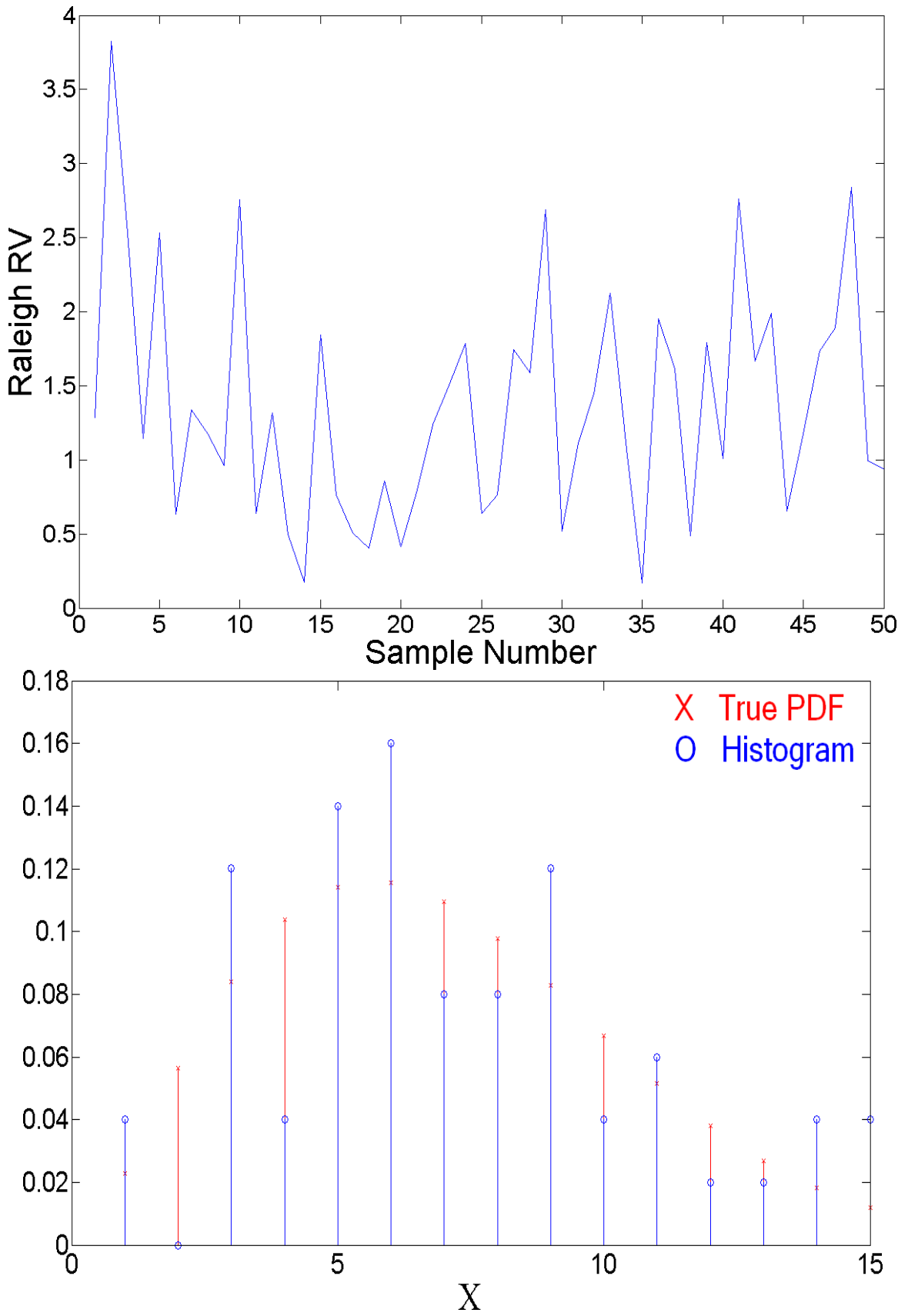

Figure-5: RV samples, estimate and base-truth PDFs, Table-1a

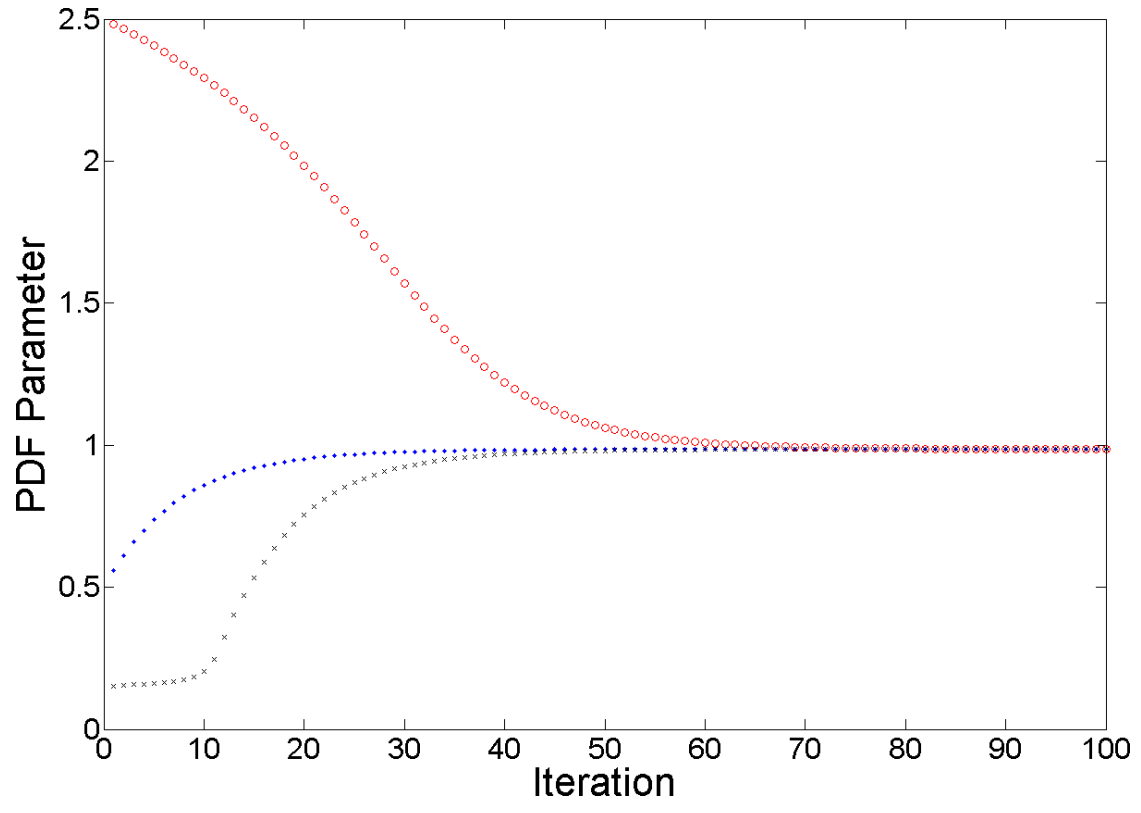

Figure-6: Estimate evolution, table-1a

The results are shown in table-1b. The Rayleigh random variables and the measurement and base-truth PDFs are shown in figure-5. The ability of the subspace approach to suppress large amount of noise is clear from the measured and base-truth PDF samples. The evolution traces for the estimates from all the initial values are shown in figure-6. All traces converged, using small number of iterations, to the same estimate. It is worth noting that changing the initial guess of the parameter has only marginal effect on the number of interactions needed to compute the estimate. Despite the low number of random variable samples used, the accuracy of the estimate is high.

| $\sigma_o$ | K | N | $\sigma_{mx}$ | $\sigma_{mn}$ |
|---|---|---|---|---|
| 4 | 50 | 15 | 10 | 0.5 |

Table-2a: Parameters of the estimator

| $\xi(0)$ | Iter | $\sigma_{est}$ | Err % |
|---|---|---|---|
| 9 | 800 | 3.279 | 18% |
| 3 | 130 | 3.277 | 18% |
| 0.7 | 240 | 3.277 | 18% |

Table-2b: Estimator results corresponding to parameters in table-2a

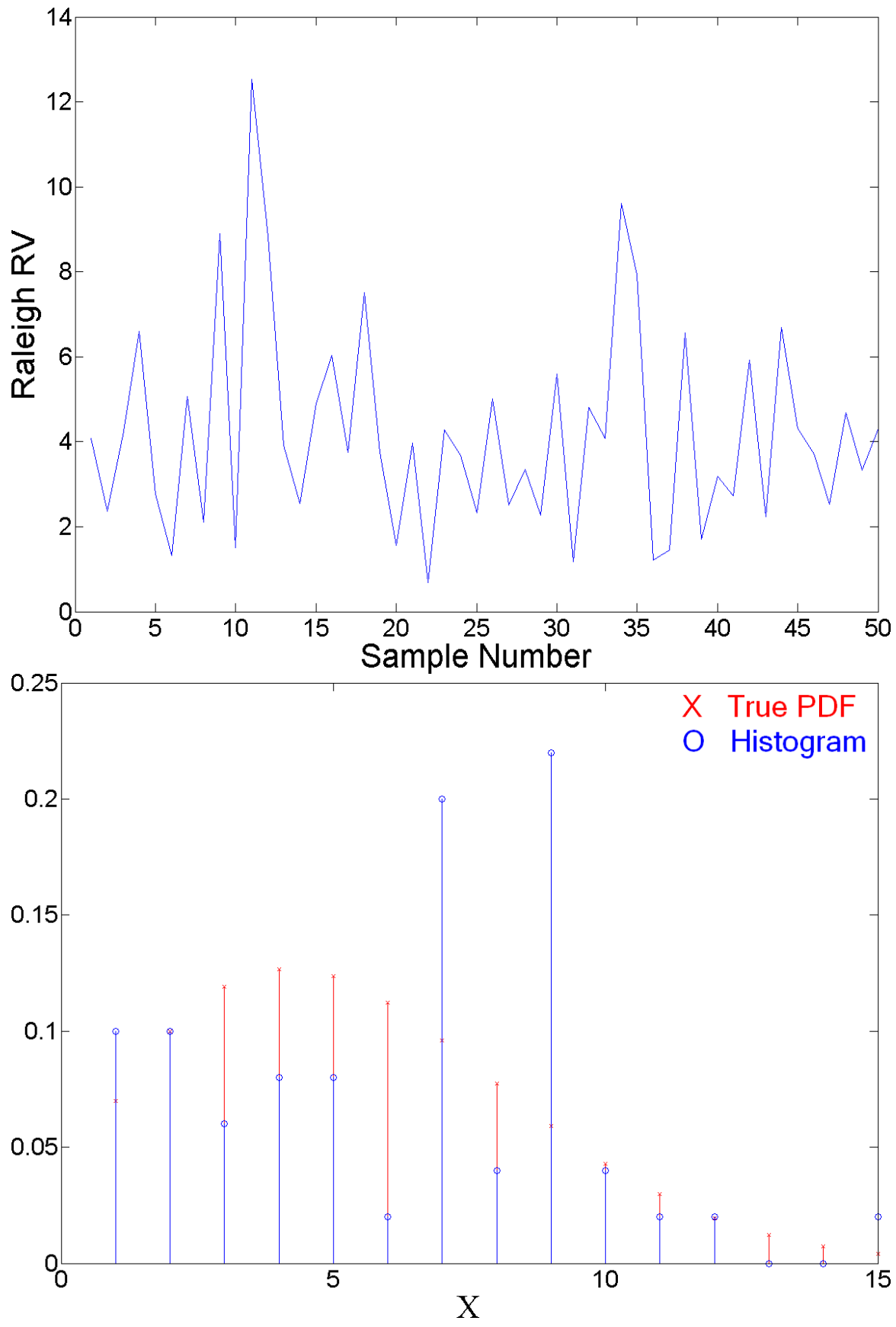

Figure-7: RV samples, estimate and base-truth PDFs, Table-2a

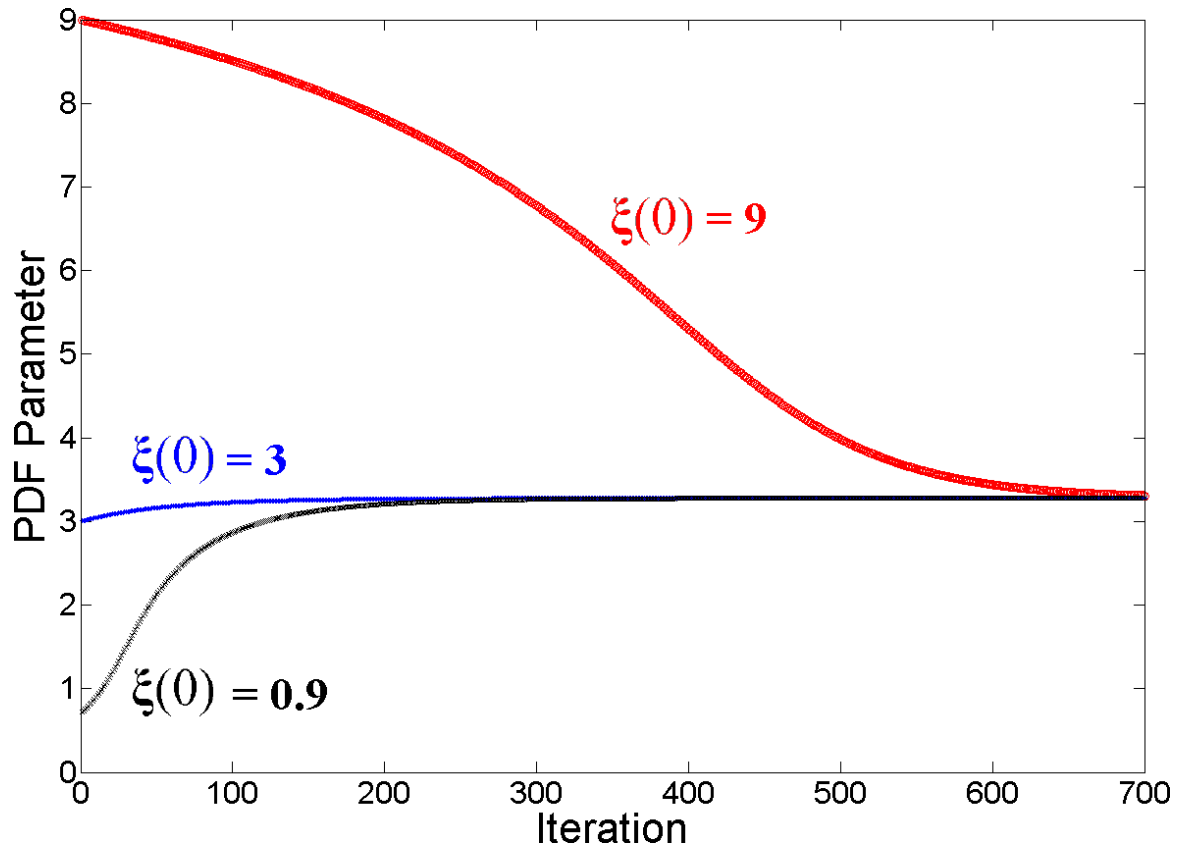

Figure-8: Estimate evolution due to different initial conditions, table-2a

| $\sigma_o$ | K | N | $\sigma_{mx}$ | $\sigma_{mn}$ |
|---|---|---|---|---|
| 1 | 500 | 15 | 2.9 | 0.17 |

Table-3a: Parameters of the estimator

| $\xi(0)$ | Iter | $\sigma_{est}$ | Err % |
|---|---|---|---|
| 2.5 | 70 | 0.998 | 0.2% |
| 1.5 | 25 | 0.998 | 0.2% |
| .3 | 25 | 0.998 | 0.2% |

Table-3b: Estimator results corresponding to parameters in table-3a

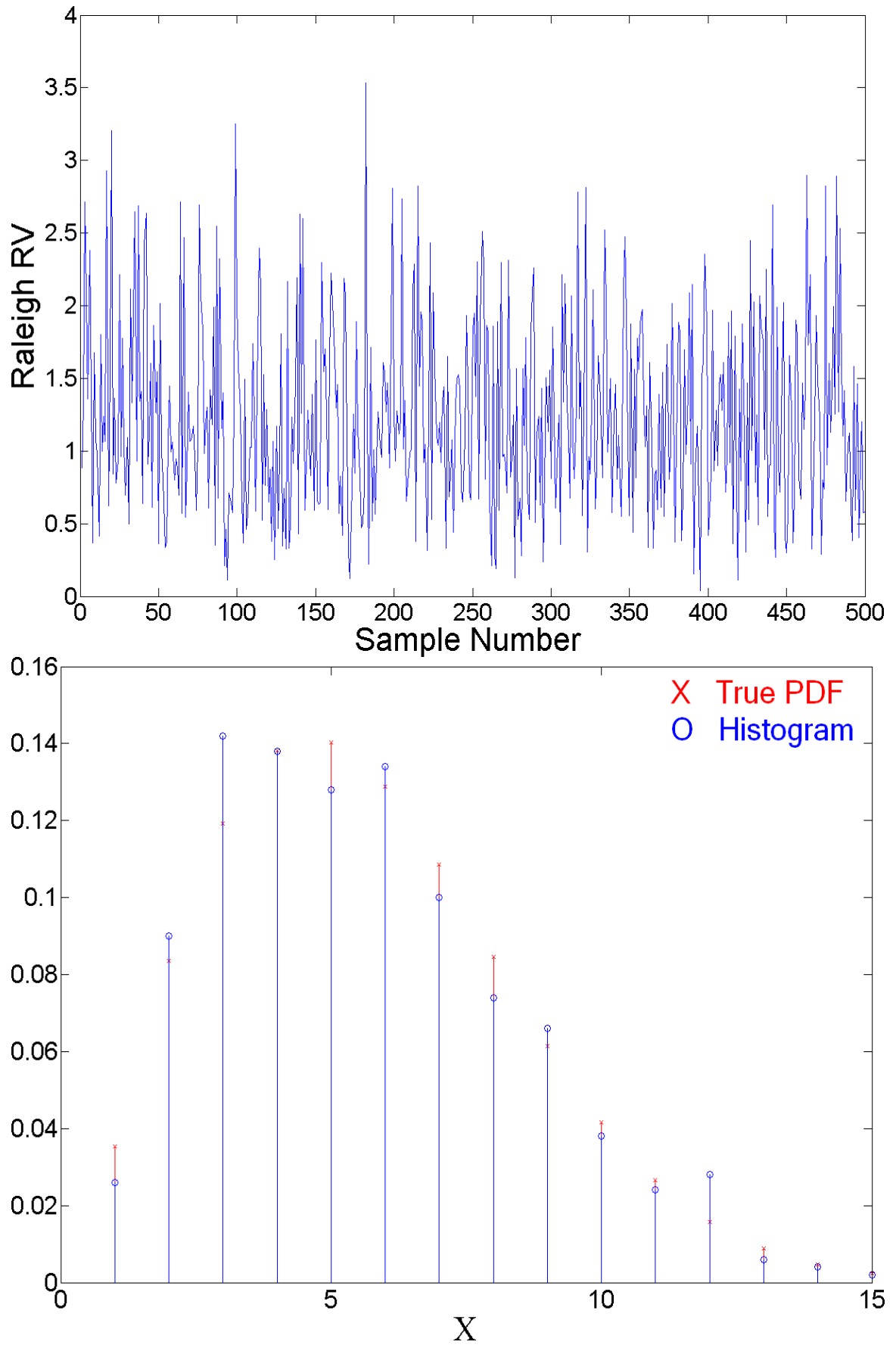

Figure-9: RV samples, estimate and base-truth PDFs, Table-3a

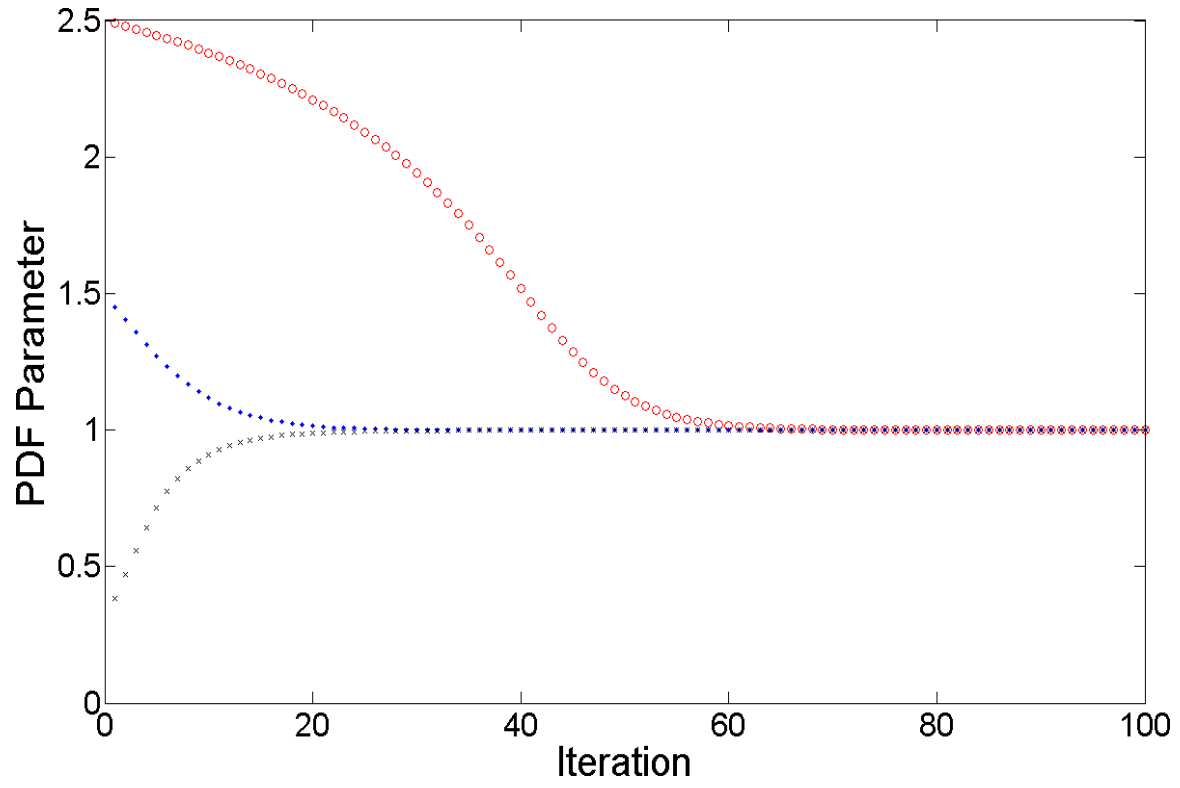

Figure-10: Estimate evolution, table-3a

The following example is similar to the previous one except for an increase in the value of the true parameter of the PDF (table-2a). As can be seen from the results (Table-2b, Figures-7,8), the characteristics of the estimator did not change, except that the speed of convergence and the accuracy deteriorated a little.

| $\sigma_o$ | K | N | $\sigma_{mx}$ | $\sigma_{mn}$ |
|---|---|---|---|---|
| 4 | 500 | 15 | 11 | 0.6 |

Table-4a: Parameters of the estimator

| $\xi(0)$ | Iter | $\sigma_{est}$ | Err % |
|---|---|---|---|
| 9 | 800 | 4.070 | 1.4% |
| 3 | 250 | 4.052 | 1.3% |
| .9 | 350 | 4.051 | 1.3% |

Table-4b: Estimator results corresponding to parameters in table-4a

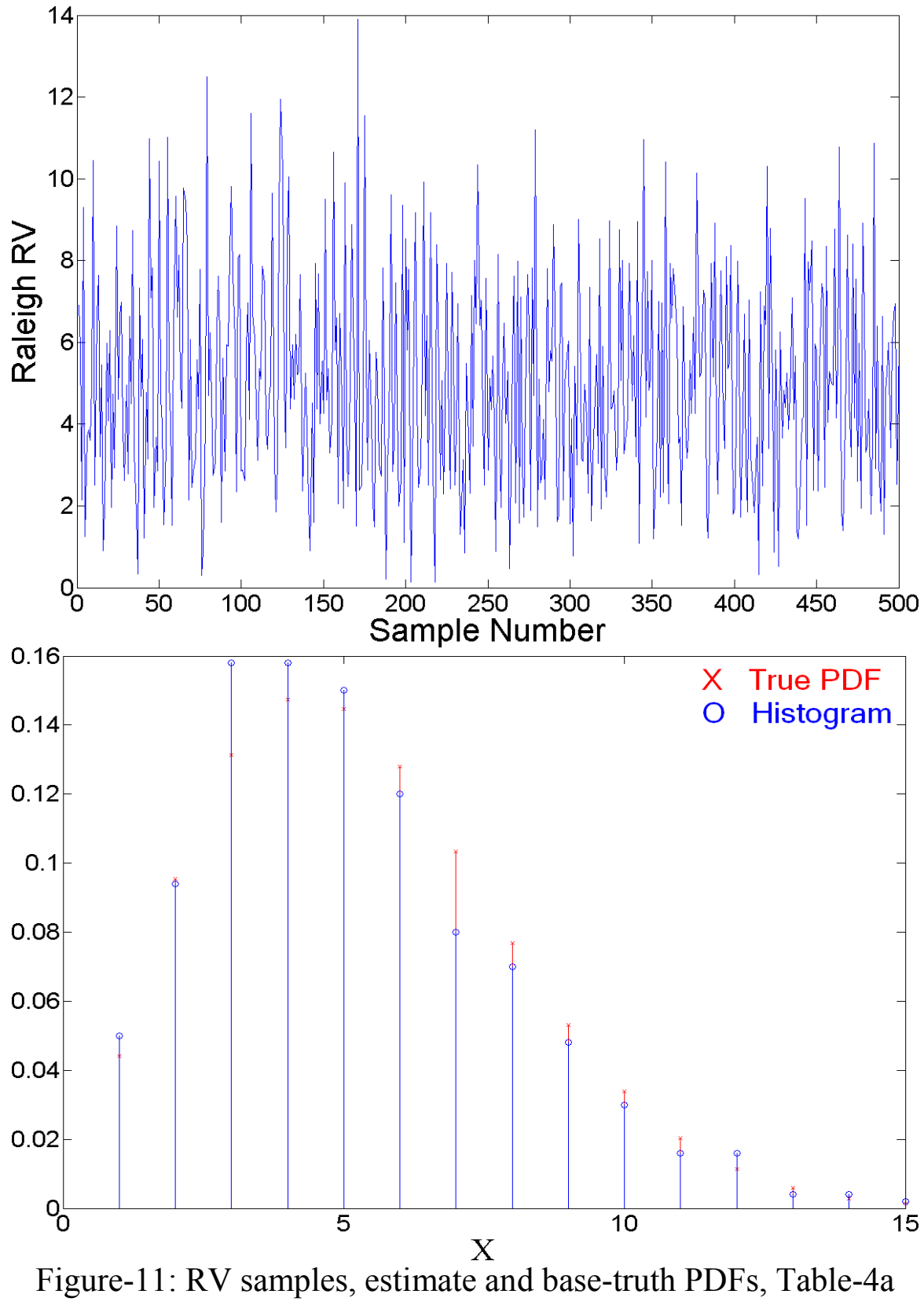

Figure-11: RV samples, estimate and base-truth PDFs, Table-4a

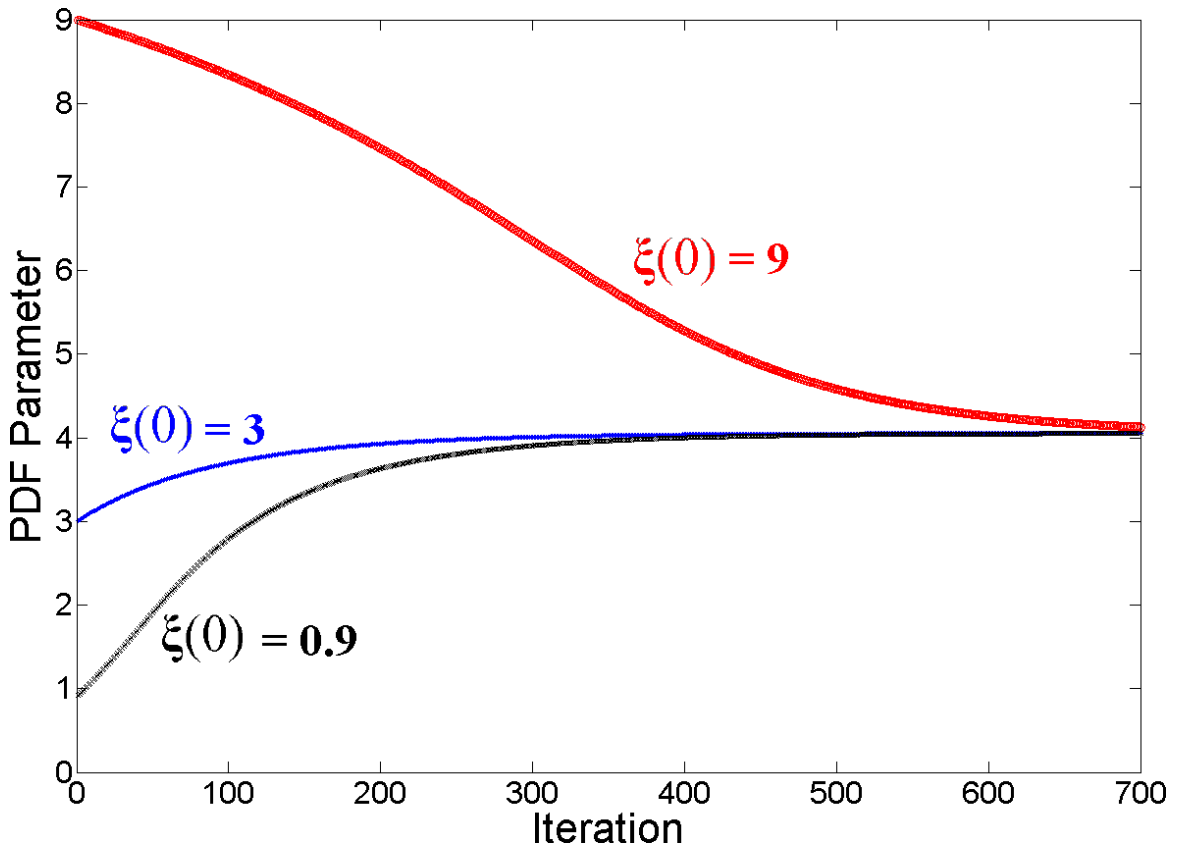

Figure-12: Estimate evolution, table-4a

To demonstrate the significant improvement in the behavior of the estimator that results from increasing the data record size, the two previous examples are repeated for $\sigma_o$=1 (Table-3a,b, Figure-9,10) and $\sigma_o$=4 (Table-4a,b, Figure-11,12) with the number of random variable samples increased to K=500. The general behavior of the estimator remained the same, the accuracy of the estimate increased, especially for higher value of the PDF parameter, and the number of iterations remained practically unaffected.

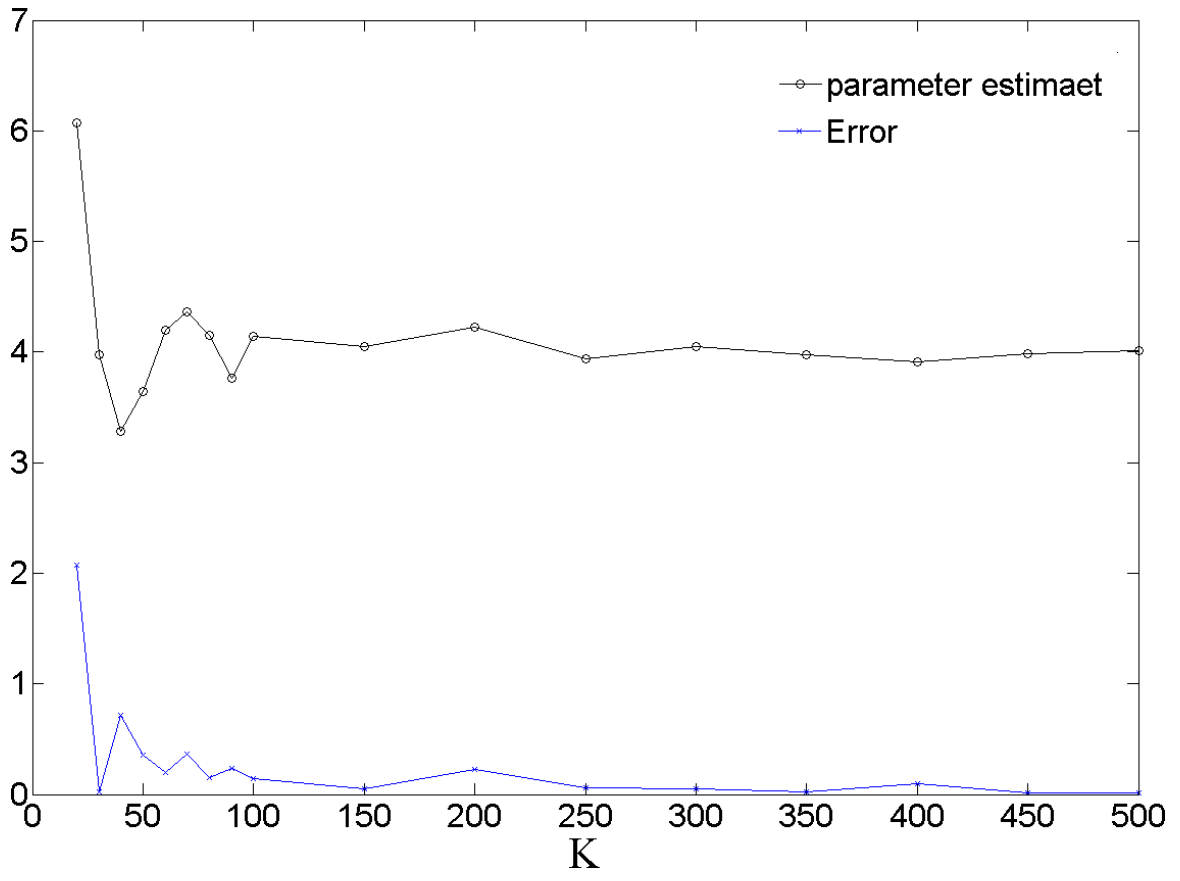

Figure-13: Estimate convergence versus number of RV samples

In figure-13 the effect of the number of random variable samples on estimate accuracy is tested for the case N=15 and $\sigma_o$=4. Rapid convergence to the correct estimate is observed as a function of K with no noticeable improvement in accuracy for K>100.

Figure-14 shows the measurement and base-truth PDFs for K=40, N=50 and $\sigma_o$=4, the high distortion in the measurement of the PDF is obvious. In figure-15 the effect of keeping K constant and varying N is tested for both K=50 and 500. It is interesting to notice the flat profile of the estimate as a function of K for both low and high number of RVs As mentioned, increasing N while keeping K constant increases the noise in the measurement. Although the fluctuation (variance) in estimate is relatively high for low number of RVs and is low for a high number of RVs, both estimate are observed to be around a constant that is equal to the true value of the estimate. This behavior could be qualitatively understood using the argument at the end of section III.

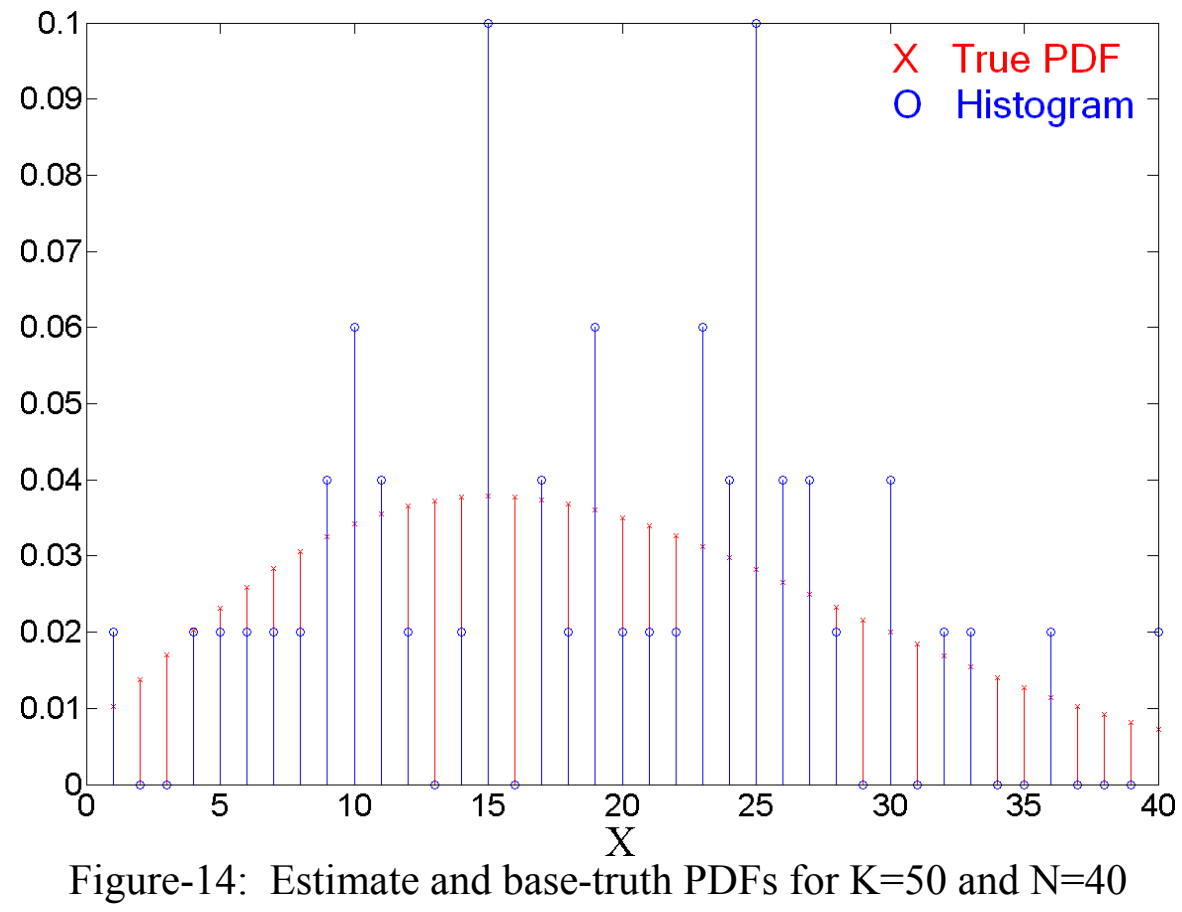

Figure-14: Estimate and base-truth PDFs for K=50 and N=40

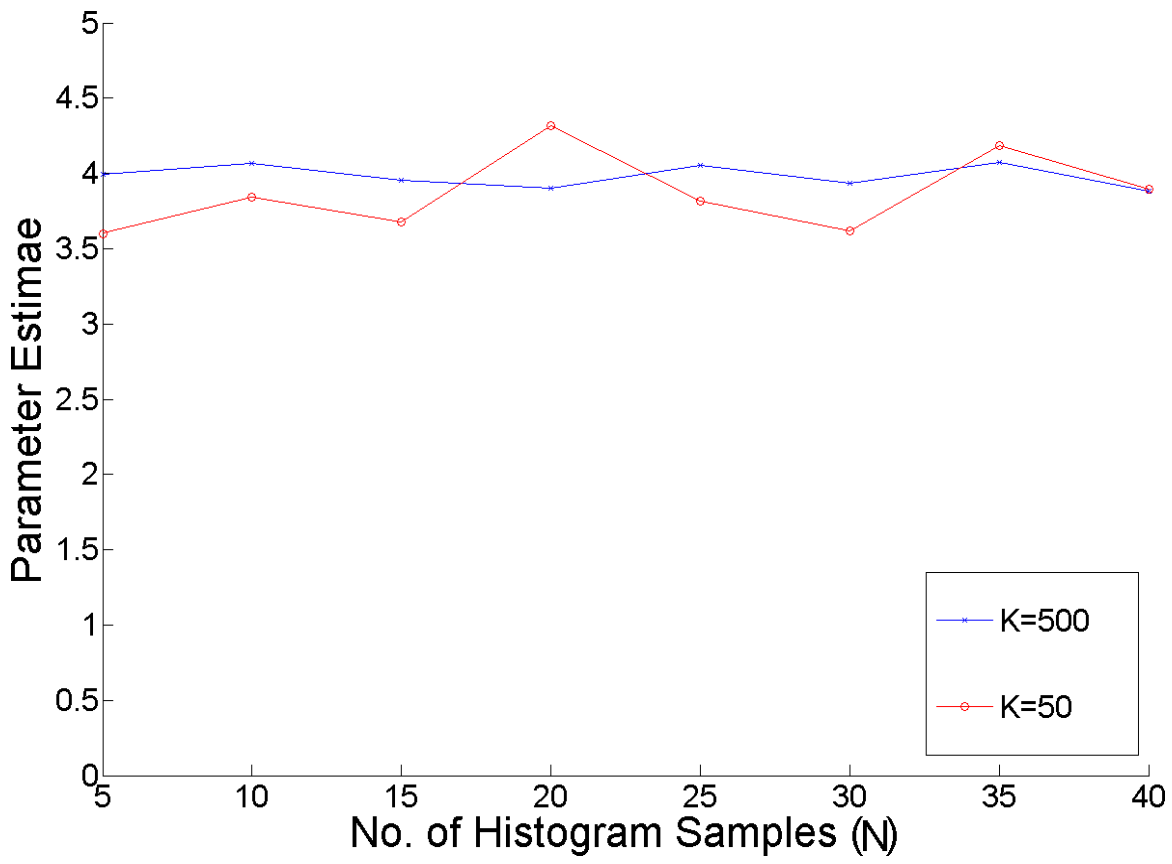

Figure-15: Estimate versus number of histogram samples (N) for constant K (K=50, K=500)

The following example clearly demonstrates the difference between obtaining the parameter estimate using the suggested nonlinear subspace method and obtaining it by direct norm minimization. In this example, 10000 records of Rayleigh random variables with parameter $\sigma$ =1 are generated. Sets of different data record lengths were used and the PDF measurements were constructed using a 15 sample histogram (N=15). Each data record was processed using the suggested method and by selecting the value of the parameter that minimizes the L2 norm:

$$L2(\sigma)=\sqrt{\sum_{i=1}^{N}(\hat{\Psi}(x_i)-\frac{x_i}{\sigma^2}\exp(-\frac{x_i^2}{2\sigma^2}))^2}\,. \qquad (18)$$

For the trials that belong to the same record length, the mean of the estimates is computed to give an indication about the estimator's bias. The variance is also computed to give an idea about the level of confidence in an estimate from a single trial representing the expected value that the estimator computes.

As can be seen from table-5 the suggested estimator outperforms direct norm minimization for all record sizes large and small. The expected value for the estimate obtained using the subspace approach remained almost constant at the true parameter value for all record size. Its variance decayed rapidly with record size. The estimates using direct norm minimization show high bias for small record size as well as high variance.

| | Subspace | | L2-Norm | |
|---|---|---|---|---|
| K | Variance | Mean | Variance | Mean |
| 30 | 0.017 | 0.9956 | .0862 | .7660 |
| 50 | .0077 | 0.9949 | .0296 | 0.9243 |
| 100 | .0029 | 0.9944 | .0092 | 0.9905 |
| 200 | .0015 | 0.9963 | .0042 | 1.0198 |
| 300 | .000973 | 1.0008 | .0030 | 1.0265 |
| 400 | .000742 | 0.9998 | .0022 | 1.0267 |
| 500 | .00057 | 1.0000 | .0017 | 1.0315 |
| 600 | .000526 | 1.0003 | .0016 | 1.0320 |

Table-5: Rayleigh parameter estimates using the suggested subspace method and by direct norm minimization (N=15).

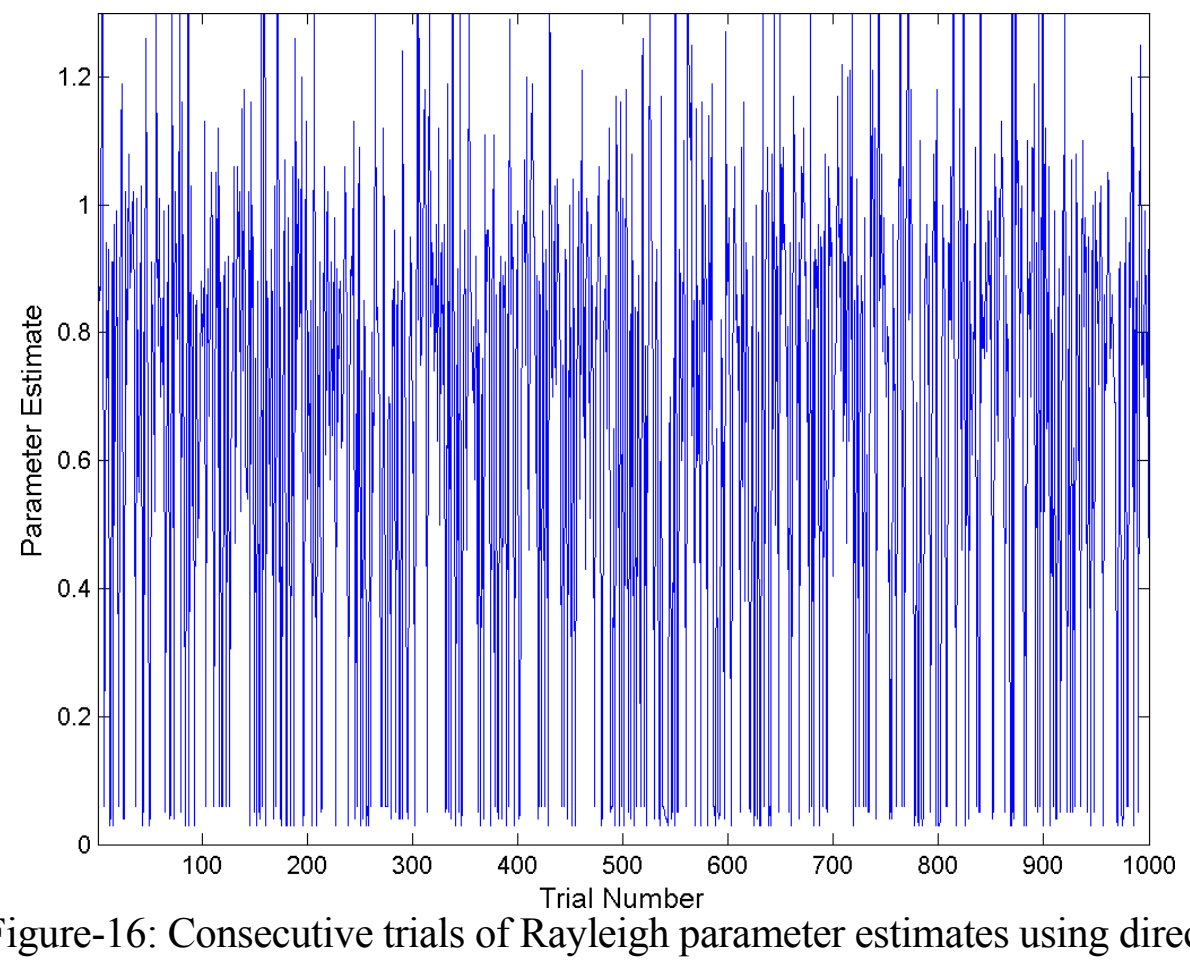

Figure-16: Consecutive trials of Rayleigh parameter estimates using direct norm minimization, K=30.

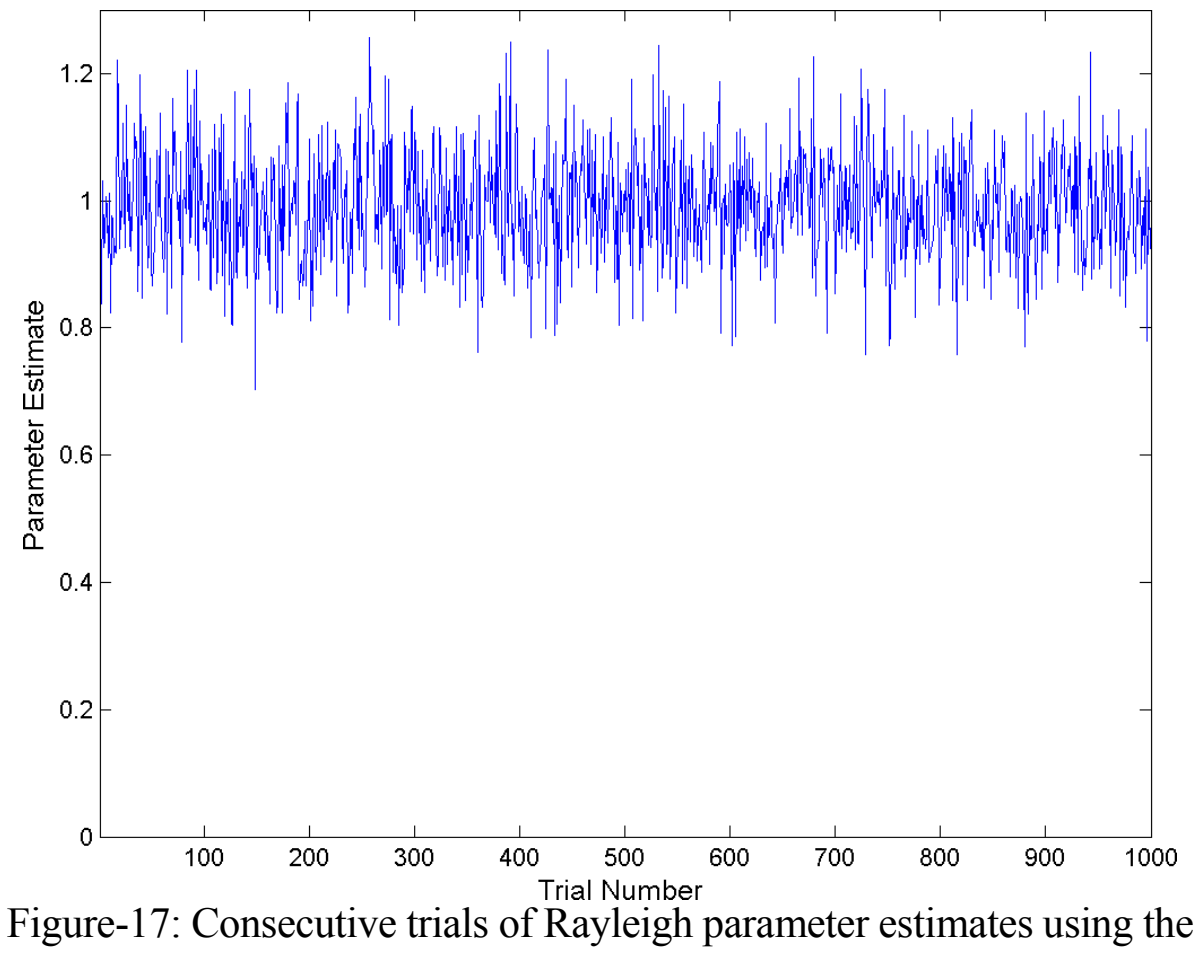

Figure-17: Consecutive trials of Rayleigh parameter estimates using the suggested subspace approach, K=30.

For small record size, the estimates obtained by direct norm minimization are practically useless. Figures 16 and 17 show the estimates from 1000 consecutive trials with record size K=30 for the direct norm minimization and subspace approach respectively. The estimates from the direct norm minimization experience rapid and large fluctuations in value. They are virtually uncorrelated with no significant influence of the expected value on the estimate. On the other hand, the estimates from the subspace method fluctuated in a narrow band around the expected value.

| | MLE | | Bayes | | Moment | |
|---|---|---|---|---|---|---|
| K | Variance | Mean | Variance | Mean | Variance | Mean |
| 30 | .0081 | 1.0004 | .0083 | 0.9871 | 0.0193 | 0.9884 |
| 50 | .0049 | 0.9987 | .0048 | 0.9913 | 0.0112 | 0.9935 |
| 100 | .0026 | 1.0016 | .0025 | 0.9941 | 0.0057 | 0.9978 |
| 200 | .0012 | 1.0003 | .0012 | 0.9976 | 0.0027 | 0.9987 |
| 300 | .00085 | 1.0007 | .00084 | 0.9990 | 0.0019 | 0.9991 |
| 400 | .00073 | 1.0003 | .00073 | 0.9994 | 0.0016 | 0.9993 |
| 500 | .00067 | 1.0004 | .00067 | 0.9996 | 0.0015 | 0.9992 |
| 600 | .00064 | 1.0002 | .00064 | 0.9994 | 0.0014 | 0.9989 |

Table-6: Parameter estimate from the data records in table-5 using MLE, Bayes and Moment methods (N=15).

In the following, the subspace method is compared to other Rayleigh parameter estimators both closed form and non-closed form. The closed form parametric estimators used are:

the maximum likelihood estimator (MLE) in [23]

$$\hat{\sigma} = \frac{4^K \cdot K!(K-1)!\sqrt{K}}{(2K)!\sqrt{\pi}} \sqrt{\frac{1}{2K}\sum_{i=1}^{K} x_i^2} \tag{19}$$

the Bayes estimator in [24]

$$\hat{\sigma} = \sqrt{K \cdot \frac{\Gamma(K+0.5)}{\Gamma(K+1.5)}} \sqrt{\frac{1}{2K}\sum_{i=1}^{K} x_i^2} \tag{20}$$

and the Moment estimator in [28]

$$\hat{\sigma} = \sqrt{\frac{\frac{1}{K-1}\sum_{i=1}^{K}\left(x_i - \frac{1}{K}\sum_{i=1}^{K} x_i\right)^2}{1-\Gamma^2(1.5)}}\,. \tag{21}$$

The non-closed form MCEM estimator in [25] is also used in the comparison.

The random variable records from the previous example in table-5 are processed using equations 19, 20 and 21 to obtain the MLE, Bayes and Moment parameter estimates. For each sample record length, the mean and the variance of the estimates from the 10000 trials are recorded in table-6. As can be seen the estimate from the suggested method almost matches the MLE and Bayes estimator and performs better than the Moment estimator.

The work in [25] reports mean estimates of the parameter for K=50 and K=100 as $\hat{\sigma} = 1.0938$ and $\hat{\sigma} = 1.03506$ respectively. As can be seen, the suggested estimator provides considerably more accurate estimates.

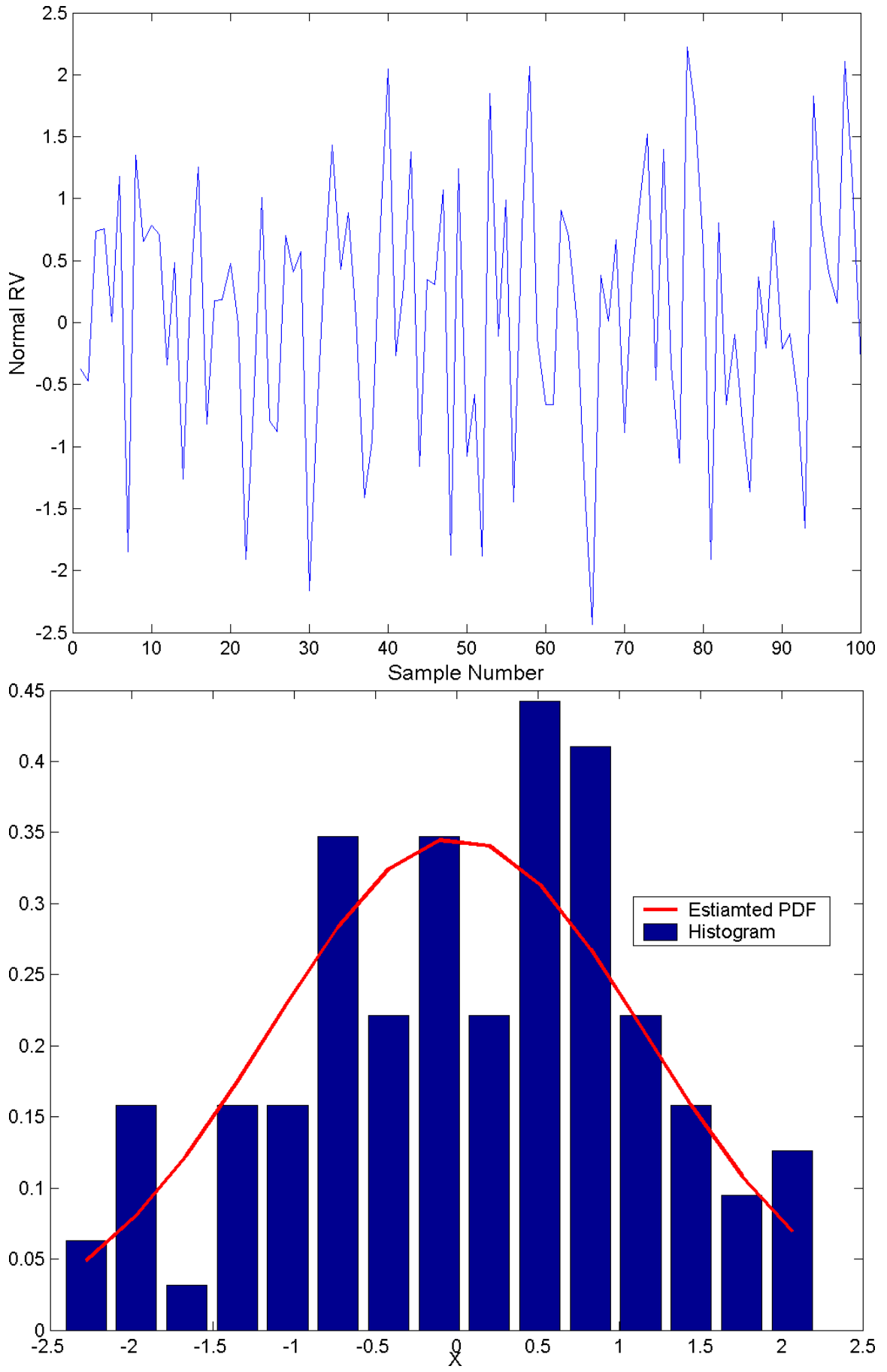

Figure-18: Normally-distributed random variables, their histogram and estimated PDF.

The following examples demonstrate the generic nature of the estimator and the ability to use the procedure for estimating PDFs other than the Rayleigh distribution. Although applying the subspace estimator to a PDF requires studying the properties of the nonlinear dynamical system that results, the estimator in its current form can handle a variety of PDF types with reasonable efficiency.

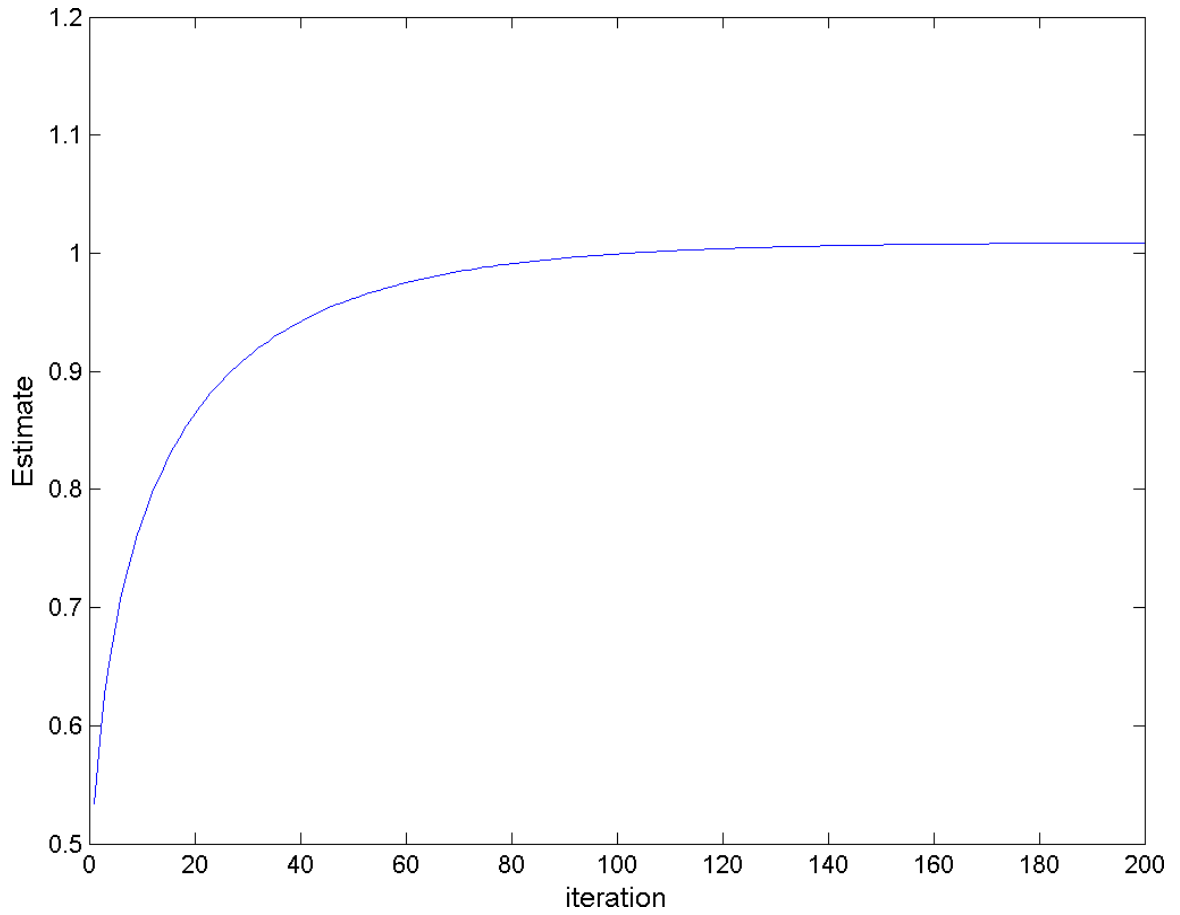

Figure-19: The evolution of normal PDF parameter estimate.

The first example (figure-18) estimates a zero-mean normally-distributed RV with σ=1 (22) from a record of 100 samples (K=100) that are used to construct a 15 sample histogram (N=15). The procedure yielded a parameter estimate σ=1.01053 in less than 200 iterations (figure-19).

$$P_X(x) = \frac{1}{\sigma\sqrt{2\pi}} \exp\left(-\frac{x^2}{2\sigma^2}\right) \tag{22}$$

The second example (figure-20) attempt to estimate the two parameters of a lognormal RV with σ=1 and μ=2 (23) from a record of 100 samples (K=100) that are used to construct a 15 sample histogram (N=15). The procedure yielded the parameter estimates σ=1.013 and μ=1.96 in less than 50 iterations (figure-22).

$$P_X(x) = \frac{1}{x\sigma\sqrt{2\pi}} \exp\left(-\frac{(\ln(x)-\mu)^2}{2\sigma^2}\right) \tag{23}$$

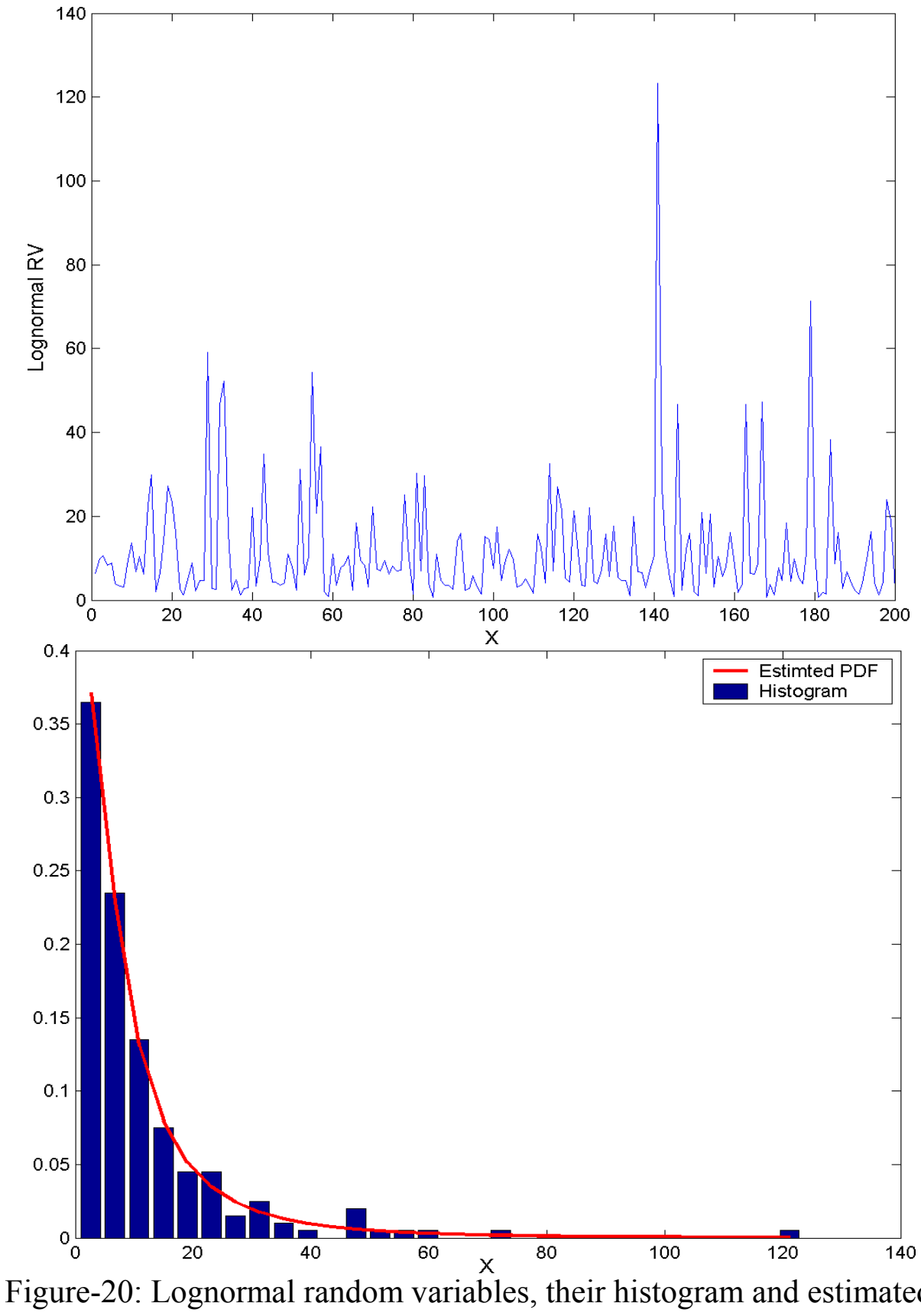

Figure-20: Lognormal random variables, their histogram and estimated PDF.

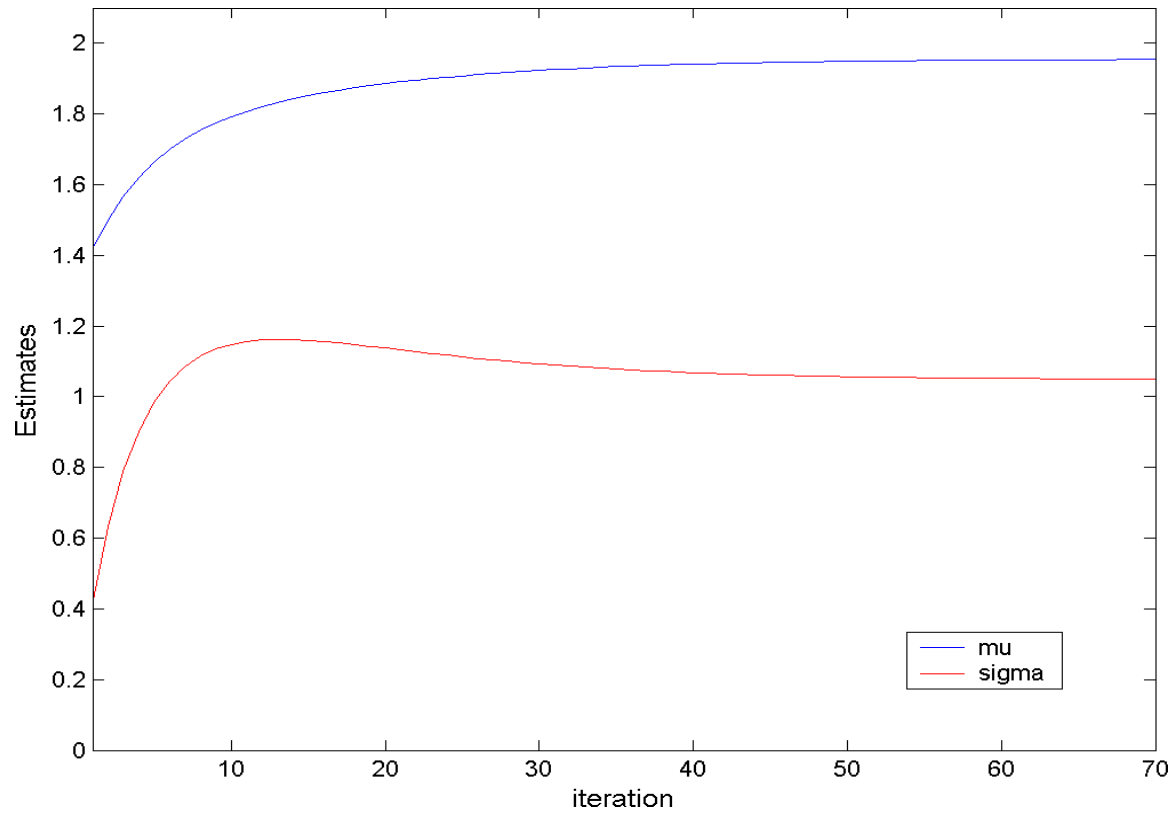

Figure-21: The evolution of the lognormal PDF parameter estimates.

## V. A NOTE ON THE ESTIMATOR

The function of the nonlinear subspace is to embed the *a priori* available information about the PDF in the estimation process. This embedding restricts the candidates on which the error norm is minimized to only those that belong to the correct type of PDF being estimated. However, the difference between the suggested nonlinear subspace estimation approach and conventional norm-based estimation techniques is more fundamental than conditional constrained estimation. It transcends that to the nature of processing being used in computing the optimum constrained estimate.

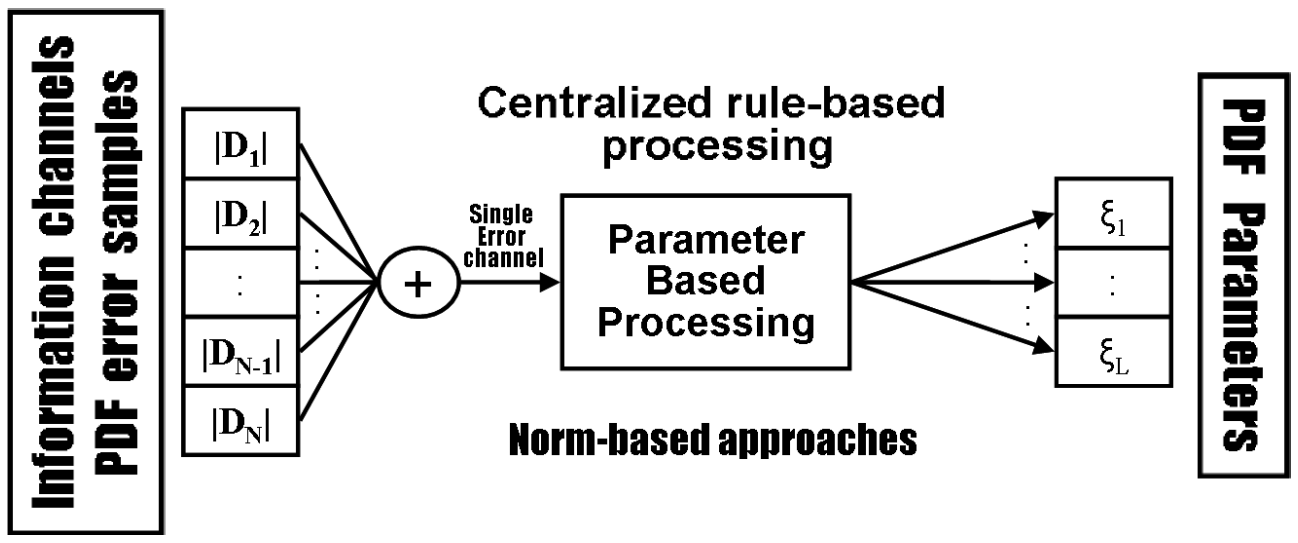

Figure-22: Structure of traditional norm-based estimators.

Norm-based methods (figure-22) attempt to reduce the complexity of the estimation process. This complexity arises from simultaneously considering the difference between the observation and estimate at each sample as a parameterized error channels whose content should be processed and used in adjusting the estimated parameters. Conventional norm-based methods average the information in all these channels to create one parameterized error whose content is to be processed to obtain the estimate. In essence, this amounts to the information-lossy process of using rough averages to discern fine structures. Added to this loss is the fact that norm-based methods ignore the sign of the error samples, which is an important source of information.

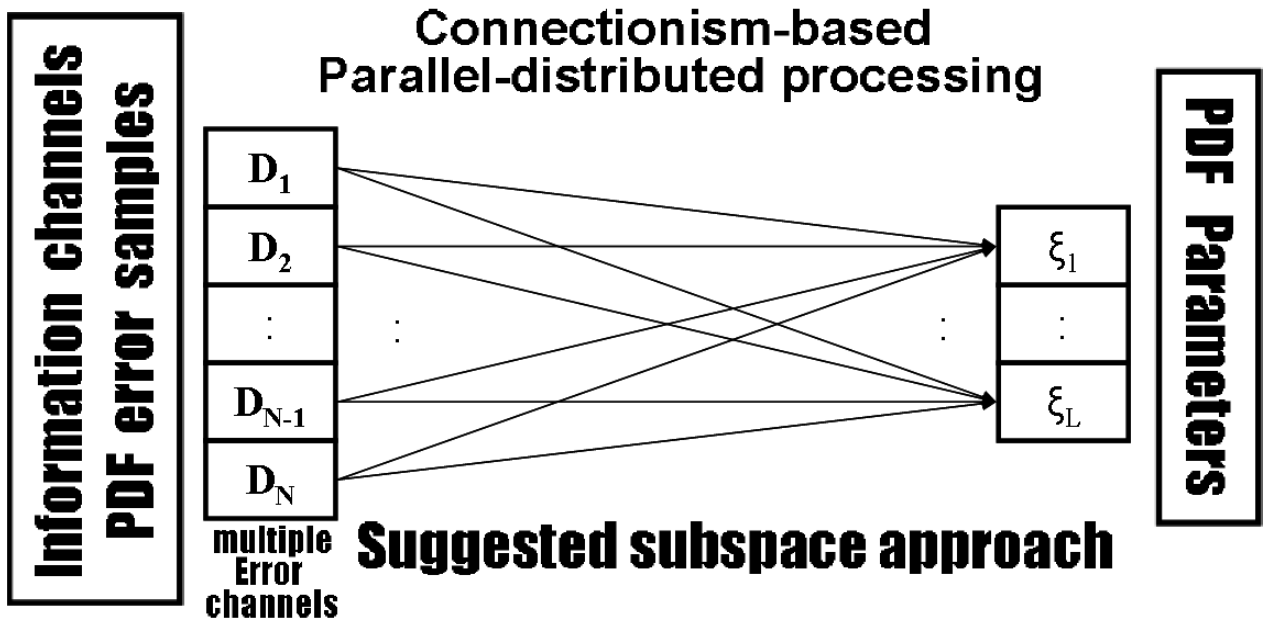

Figure-23: Parallel-distributed structure of the subspace estimator.

The suggested nonlinear subspace method processes each error sample individually. It enables the outcome of this processing collaboratively and directly to form the value of the estimate. Even a single error sample is capable of producing an estimate. In essence, the subspace processor is a complex dynamical system that employs massive parallel-distributed processing in generating the estimate. Parallel-distributed processors are hardware friendly and can perform in real-time. Software implementation, as demonstrated by simulation, is efficient. The number of multiplications and additions needed to generate an estimate at an iteration is linear in the number of samples and the number of parameters (Number of additions=Number of Multiplications = N·L).

## VI. CONCLUSIONS

This paper suggests a novel and efficient subspace-based method for the parametric estimation of the Rayleigh PDF from short data records. The principle on which the method operates is developed, proof of its correctness is provided along with simulation results to demonstrate its capabilities. The method, in principle, may be used to estimate any PDF. However, the emphasis here is on the Rayleigh PDF whose estimation from short data record is important in communication-aware mobility. The work also shows that attempting to find parameter estimates by direct norm minimization is, at least, not practical. The concept of information-based constrained estimation produces more reliable results.

Although future work will focus on theoretical analysis of the estimator's properties, the method, in its current form, may be used in an efficient and practical manner in Rayleigh PDF estimation.

**Acknowledgement:** The author would like to thank King Fahad University of petroleum and minerals for its support of this work. The author is a member of the center for communication systems and sensing.

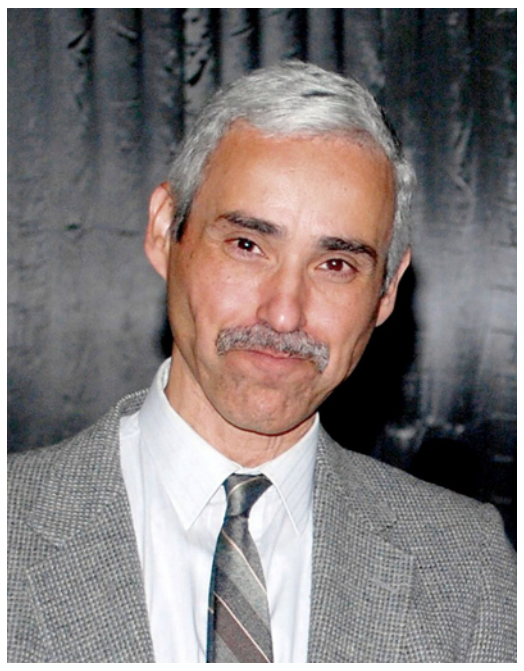

Ahmad A. Masoud: Received the B.Sc. degree in electrical engineering with a major in power systems and a minor in communication systems from Yarmouk University, Irbid, Jordan, in 1985, and the M.Sc. degree in signal processing and the Ph.D. degree in robotics and autonomous systems from the Electrical Engineering Department, Queen's University, Kingston, ON, Canada, in 1989 and 1995, respectively. He worked as a Researcher with the Electrical Engineering Department, Jordan University of Science and Technology, Irbid, from 1985 to 1987. He was also a part-time Assistant Professor and also a Research Fellow with the Electrical Engineering Department, Royal Military College of Canada, Kingston, from 1996 to 1998. During that time, he carried out research in digital signal processing-based demodulator design for high density, multiuser satellite systems, and taught courses in robotics and control systems. He is currently an Assistant Professor with the Electrical Engineering Department, King Fahd University of Petroleum and Minerals, Dhahran, Saudi Arabia. His current interests include navigation and motion planning, robotics, constrained motion control, intelligent control, and DSP applications in machine vision and communication.